\documentclass[journal,onecolumn,twoside]{IEEEtran}

\ifCLASSINFOpdf
\else
\fi
\usepackage{diagbox} 
\usepackage{algorithmic}
\usepackage{algorithm}

\newtheorem{theorem}{Theorem}
\newtheorem{lemma}{Lemma}

\usepackage{cite}

\usepackage{graphicx}  
\usepackage{epsfig}
\usepackage{amsmath}
\usepackage{booktabs}

\usepackage{amssymb} 
\usepackage{setspace}
\begin{document}

%
\title{Sampling Graphlets of Multi-layer Networks: A Restricted Random Walk Approach}
%
%
%

\author{Simiao~Jiao, Zihui~Xue, Xiaowei~Chen, 
        and~Yuedong~Xu
\thanks{Simiao Jiao, Zihui Xue and Yuedong Xu are with School of Information Science and Engineering. Email: smjiao@outlook.com, sherryxue9@gmail.com, ydxu@fudan.edu.cn}
\thanks{Xiaowei Chen is with Baidu Inc. Email: veralchen@gmail.com}
}

%
%

\markboth{Journal of \LaTeX\ Class Files,~Vol.~14, No.~8, August~2015}%
{Shell \MakeLowercase{\textit{et al.}}: Bare Demo of IEEEtran.cls for IEEE Journals}
%



\maketitle

\begin{abstract}
Graphlets are induced subgraph patterns that are crucial to the understanding of the structure and 
function of a large network. A lot of efforts have been devoted to calculating graphlet statistics where 
random walk based approaches are commonly used to access restricted graphs through the available application 
programming interfaces (APIs). However, most of them merely consider individual networks while overlooking the strong 
coupling between different networks. In this paper, we estimate the graphlet concentration in 
\emph{multi-layer} networks with real-world applications. An inter-layer edge connects two nodes in different layers if they 
belong to the same person. The access to a multi-layer network is restrictive in the sense that 
the upper layer allows random walk sampling, whereas the nodes of lower layers 
can be accessed only though the inter-layer edges and only support random node or edge sampling.  
To cope with this new challenge, we define a suit of two-layer graphlets and propose novel random walk sampling algorithms to estimate the proportion of all the 3-node graphlets. An analytical bound on the sampling steps is proved to guarantee the convergence of our unbiased estimator. We further generalize our algorithm to explore 
the tradeoff between the estimated accuracies of different graphlets when the sample budget is split on different layers. 
Experimental evaluation on real-world and synthetic multi-layer networks demonstrate the accuracy and high efficiency of our unbiased estimators.
\end{abstract}


%
\IEEEpeerreviewmaketitle

\section{Introduction}
Complex networks have attracted great attention due to their ample examples in real-world such as road networks \cite{roadnetwork,roadnetwork2}, social networks \cite{socialnetwork,socialnetwork2}, and biological networks \cite{biologicalnetwork}. 
With an enormous amount of data in these fields being available, significant advances in understanding 
the structure and function of networks, and mathematical models of networks have been achieved in the 
past decade. Extensive efforts have been devoted to characterizing network properties, including 
measures of degree distribution, node clustering, network modularity, local graph structures and so on \cite{biologicalnetwork,nodeclustering,networkmodularity}. 
However, the literature deals almost exclusively with \textit{single-layer} networks whose nodes 
and edges of a network exist in an isolated system. In many state-of-the-art systems, an individual network 
is actually one component within a more complicated \emph{multi-layer} network, or shows  a strong coupling 
with other networks. Consider the scenario where we have two Online Social Networks (OSNs) 
Facebook and Twitter. Because of the diversity in their services, a fraction of users may possess the identities of both sites, 
thus linking them together. 
We define there exists a link between two accounts in Facebook and Twitter respectively if they belong to the same person. 
A single-layer model is not capable of capturing the links between these networks. Multiple-layer (two-layer by default) networks which consist of layers of several networks overcome the disadvantages of single-layer models and are able to measure interactions across different networks. Still take Facebook and Twitter as examples. 
These two OSNs can form a two-layer network with 
each OSN on a layer. The ``\textit{intra-layer}'' links (links within one layer) are friend relationship between accounts in the corresponding OSNs, and the ``\textit{inter-layer}'' links are connections between the same person in Facebook and Twitter. More examples on multi-layer networks 
include the cyber-physical systems where one layer can be a physical acquaintance network of users.

Graphlets, which are referred to as induced subgraph patterns or motifs, are the building block of complex networks. 

One famous example in the graphlet family is the triangle. 
Computing graphlet counts in a network is an important task because the 
frequencies of graphlets offer important statistics to characterize the local topology structures.
For instance, Heider developed the balance theory \cite{heider1958psychology} that uses 3-node graphlets to explain social proverbs - ``A friend of my friend is my friend'' and 
``The enemy of my enemy is my friend''
. Kunegis \emph{et al.} took the concentration of graphlets as a metric to gauge the stability of signed friend or foe 
subgraphs \cite{kunegis2009slashdot}. Juszczyszyn \emph{et al.} used the triad transition pattern to predict whether a link would be constructed 
between pairwise users at a future time. Rahman and Hasan proposed to extract feature representation of graphlet transition 
events for link prediction in dynamic networks \cite{juszczyszyn2011link}. 

\begin{figure}[h]
  \centering
  \includegraphics[width=4in]{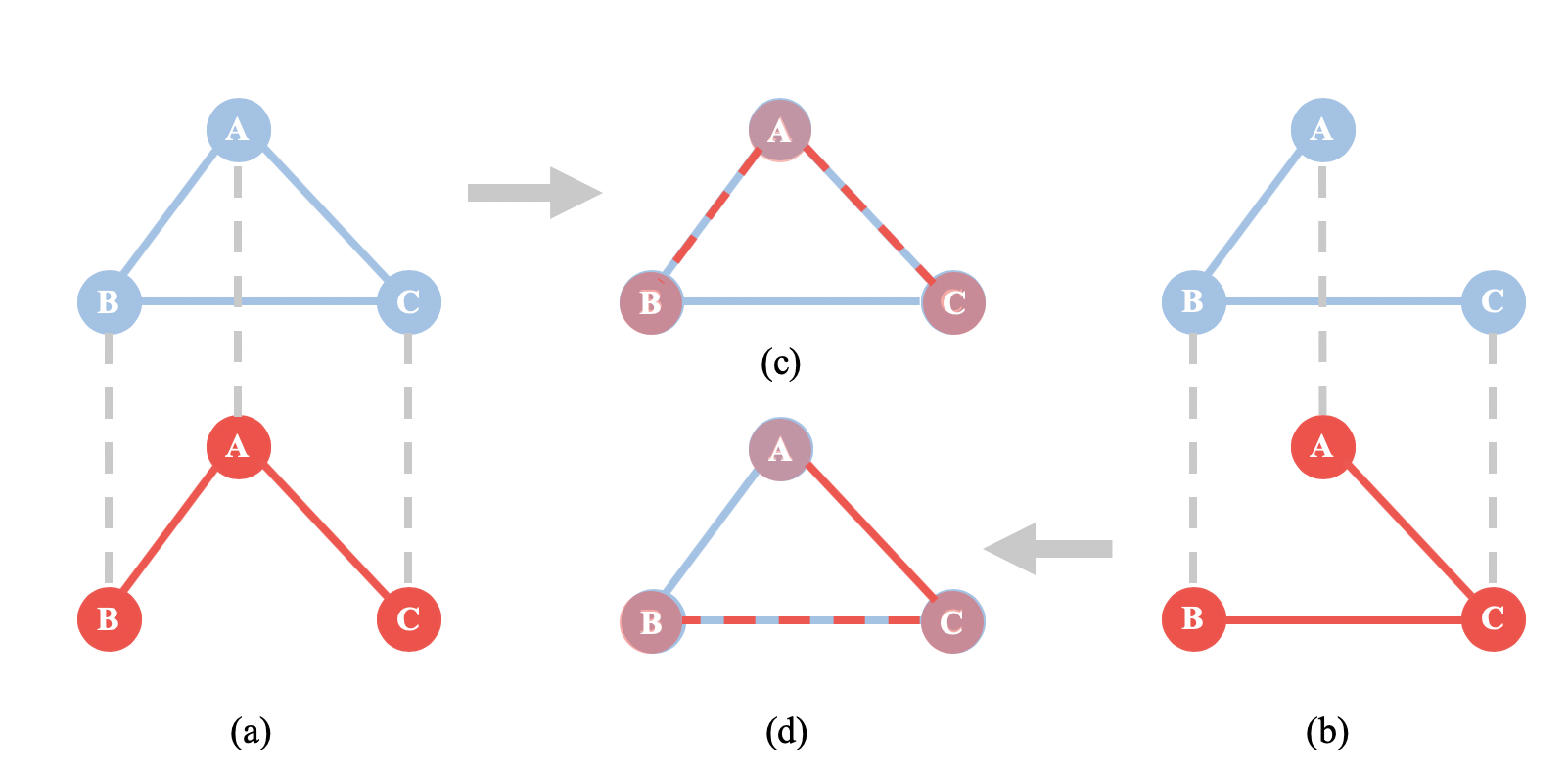}
  \caption{\textbf{An illustration of the two-layer network}}
\label{fig:two_layer}
\end{figure}

Despite of the comprehensive research on graphlets in single-layer networks, it remains largely open in multi-layer networks. 
The definition of multi-layer graphlets itself is the first obstacle in which we generalize the single-layer counterparts to 
this new scenario. Taking the two-layer network in Fig. \ref{fig:two_layer} as a simple example where the upper-layer edges 
are colored in \emph{blue}, and the lower-layer ones are colored in \emph{red}. The black dashed lines indicate that the corresponding nodes appear in both layers. The subgraphs in Fig.\ref{fig:two_layer} (a) and (b) represent the two-layer 3-node
graphlets that characterize their local structures in the blue and red graphs. 
One can squish a two-layer graphlet into a single-layer subgraph, e.g. Fig.\ref{fig:two_layer} (a) to (c) and Fig.\ref{fig:two_layer} (b) to (d), and use 
different colors to differentiate the edges in different layers. 
If not mentioned explicitly, we choose the squished subgraphs (e.g. Fig \ref{fig:two_layer} (c) and (d)) to visualize the two-layer graphlets for simplicity throughout this work. The multi-layer graphlets not only inherit their significance in each individual networks, but also 
reveal the interactions between networks. The graphlet in Fig. \ref{fig:two_layer} (a) manifests a strong tie among three users due to their 
dense connectivity in both layers. In Fig. \ref{fig:two_layer} (b), user A can 
expand the coverage of service recommendation: diffusing an information to his 
neighbor B at the blue layer and reaching user C at the red layer. 
Furthermore, the triad based friendship recommendation \cite{zhang2015diffusion} and 
spam detection \cite{becchetti2008efficient} can benefit from the multi-layer structure.

The purpose of our study is to efficiently compute the frequencies that each 
two-layer graphlet appear in a given graph. The percentage of a particular graphlet type is called 
the ``graphlet concentration'' or ``graphlet statistics''.

\textbf{Challenges.} The first challenge is the prohibitive complexity of exhaustive counting. As is well known, there exist a huge amount of graphlets even for a moderately sized single-layer graph.For two-layer networks, the time complexity of
exhaustive counting is even higher because there are more distinct graphlets. 

Therefore, we resort to an alternative approach 
that ``samples'' a small fraction of nodes and edges of the graph in order 
to significantly reduce running time with an acceptable error level. 
A string of sampling algorithms have been proposed to estimate graphlet concentration so far \cite{memorybased1,memorybased2,memorybased3,memorybased4,streaming1,streaming2,restrict-cluster,restrict-graphlet,restrict,restrict-chen,reservoirsample,ahmed2017sampling}. 

The second challenge is the restricted access to the complete graph data. The restriction is even more stringent in 
multi-layer networks than in single-layer ones which gives rise to a fundamentally different rationale in designing sampling 
approaches. The OSN service provides are unwilling to share the complete graph, but allow the calling of some application programming interfaces (APIs). With these APIs, we can query a node and retrieve a list of the neighboring edges. Random 
walk \cite{randomwalk} is the \emph{de facto} approach to tackle the restricted graph access problem in single-layer networks. 
However, \emph{additional restrictions} have been imposed on the sampling in two-layer networks. Consider a ``social-hub'' network \cite{gong2019exploring} consisting of 
a Twitter-like open social network 
(upper layer) and a Facebook-like privacy sensitive social network (lower layer)\footnote{The default privacy setting of existing OSNs usually goes to two extremes. In Twitter, a stranger is allowed to visit the friend list of a user and explores the friends of his friends while in LinkedIn, a stranger is forbidden to see the friend list. 
We consider a semi-private setting that will hopefully be considered in the future: a stranger is allowed to see 
the friend list of a user, but is refused to explore the 
friends of his friends.}. The upper layer allows random walks through the provided APIs, while the lower layer 
only lists the direct friends of a sampled user. 
In a coupled cyber-physical network with a Facebook layer and a physical acquaintance layer. The crawler can randomly walk on the Facebook graph, but it is costly to 
dive into the physical layer to ``query'' all the real-world friends of a person or his friends-of-friends. 

\emph{In a word, the sampling of two-layer graphlets allows random walk on one layer and only node or edge 
sampling on the other layer.} Such a restriction makes the existing single-layer random walk approaches no longer applicable. 

\subsection{Related work} 
\noindent\textbf{Exact Graphlet Counting.} 
Counting subgraphs is a computationally intensive task. 
Ahmed \emph{et al.} proposed a fast parallel algorithm that leverages a number of proven combinatorial arguments for 
different graphlets with 3 or 4 nodes \cite{ahmed2015efficient}. Hocevar and  Demsar \cite{hovcevar2014combinatorial} proposed a combinatorial method that 
builds a system of equations to allow the computing of the number of occurrence of graphlets with up to five nodes. 
Suri \emph{et al.} \cite{suri2011counting} presented a triangle counting algorithm tailored for MapReduce computation in parallel. 

\noindent\textbf{Single-layer Network Sampling.} Existing sampling methods can be classified into three categories 
according to the graph access modes. The first one is memory-based, in which the entire graph can be directly accessed 
so that either the node sampling or the edge sampling can be easily implemented \cite{memorybased1,memorybased2,memorybased3,memorybased4}. 
The second one is the graph stream. A streaming graph is a rapid, continuous, and possibly unbounded time-varying stream of edges \cite{streaminggraph} that is usually hard to be put in a small memory or is not fully observable at any time. 
Some related works used a traversal on the edge streaming, e.g. \cite{streaming1}, and some others 
adopted sampling, e.g. \cite{streaming2}. 
Single-pass reservoir sampling schemes are proposed in \cite{reservoirsample} in which the sample is maintained incrementally over the stream, and remains useful at any time for stream property estimations. Ahmed \emph{et al.} \cite{ahmed2017sampling} proposed the
graph priority sampling method, an improvement over order-based reservoir sampling, through weighting the sampling of edges.
The third category is the restricted access. For a large network, we can only access it by using 
some application programming interfaces (APIs), for example, large OSNs such as Facebook or Twitter. 
To sample this graph, we need to crawl the identity of a person (i.e. a node) as well as his neighbours, and then randomly 
select one of his neighbours to crawl repeatedly. This method is named random walk, and is commonly used to 
estimate the degree distribution\cite{restrict-other,restric-other2}, the concentration of graphlets \cite{restrict-chen,restrict-graphlet,restrict} or the clustering coefficient \cite{restrict-cluster}.

\noindent\textbf{Multi-layer Network Sampling.}
The sampling of multi-layer heterogeneous networks is very rarely studied. 
The plausible reason is that the classical sampling algorithms 
can be generalized to the situations where the nodes or the edges are of multiple properties. 
Gjoka \emph{et al.} observed multi-type relationships (edges) among nodes of OSNs, combine all individual 
graphs into a single union graph before sampling \cite{gjoka2011multigraph}. 
Li and Yeh decomposed an OSN into a multiple-layer one in which the nodes at the same layer possess the same identity \cite{li2011sampling}. 
Three known methods (i.e. random node, random walk and respondent-driven sampling) are applied to estimate the node or edge type distributions.
To the best of our knowledge, there is no earlier study on the random walk sampling of multi-layer networks where 
the graph is not fully accessible and the visit of one layer is feasible upon which the associated nodes of the other layer 
have been visited before. Especially, it remains untouched when the sampling methods on multiple layers are different and coupled. 

\subsection{Our Contributions}

\textbf{Novel Sampling Problem.} The novelty of this work lies in three aspects. 
To the best of our knowledge, we are the first to investigate the restricted sampling problem in multi-layer networks 
that has ample real-world examples. On the upper layer, the access to the complete raw data is prohibited, but the indirect 
access through APIs is allowed. On the lower layer, the access to a node's neighboring nodes or edges is possible on 
when the corresponding node in the upper layer has been visited. This coupled restriction has not been considered 
previously, and makes the graphlet sampling very tricky meanwhile. 
Secondly, we design a novel sampling algorithm that uses the visited Markov 
states to infer the 3-node graphlet concentration with 
isomorphic state precomputation. Thirdly, for a given budget of sampling steps, 
we explore the tradeoff between the accuracies of different graphlets when more nodes or edges are sampled 
at the lower layer network. Owing to heterogeneous time or economic cost of sampling an edge in different layers, appropriately 
assigning the sampling budget may benefit the accuracy of specific types of graphlets. 
We believe that the raised problem will elicit a good many 
works in new scenario and new sampling algorithms in multi-layer networks.

\noindent\textbf{Provable Guarantee.} We prove the unbiasedness of our sampling algorithms, and derive an 
analytic Chernoff-Hoeffding bound on the needed sample size to achieve a certain error. Especially, the theoretical 
analysis shows how the performance of sampling is influenced by network parameters.

\noindent\textbf{Extensive Experimental Analysis.} Extensive experiments on real-world and synthetic multi-layer networks 
are conducted to evaluate the accuracy of graphlet concentration. Experimental results confirmed the unbiasedness, accuracy and convergence of the proposed estimators. Our algorithms demonstrate comparable accuracy with the random walk sampling on both layers with no restriction.

The remainder of this paper is organized as follows. Section \ref{sec:formulation} presents the problem formulation. 
We design a novel sampling algorithm with provable performance in Section \ref{sec:algorithm}. Section \ref{sec:extension}  explores the tradeoff of assigning sampling steps in different layers. 
Section \ref{sec:experiment} evaluates the proposed algorithm on synthetic and real-world multi-layer networks and 
Section \ref{sec:conclusion} concludes this work.

\section{Problem Formulation}\label{sec:formulation}

In this section, we present the multi-layer network model and the theoretic 
foundations of graph sampling. 

\subsection{Network Model}

A connected complex network is denoted by $\mathcal{G} = (\mathcal{V}, \mathcal{E})$ where 
$\mathcal{V}$ is the set of nodes and $\mathcal{E}$ is the set of undirected edges. 
We need multiple colors to characterize the relationship of pairwise nodes on multiple layers. Denote by $\mathcal{V}_B \subseteq \mathcal{V}$  the set of first layer nodes colored in \textbf{BLUE} 
and $\mathcal{V}_R \subseteq \mathcal{V}$ the set of second layer nodes colored in \textbf{RED}. 
Similarly, let $\mathcal{E}:=\{\mathcal{E}_B, \mathcal{E}_R, \mathcal{E}_C\}$ where $\mathcal{E}_B$ and $\mathcal{E}_R$ 
are the set of edges in the blue and red graphs respectively, 
and $\mathcal{E}_C$ consists of the edges connecting two graphs. An edge in $\mathcal{E}_C$ means that the two nodes 
belong to the same person. Let $\mathcal{G}_B = (\mathcal{V}_B, \mathcal{E}_B)$ and 
$\mathcal{G}_R = (\mathcal{V}_R, \mathcal{E}_R)$ be the corresponding blue and red graphs. 
There have been a variety of real-world counterparts regarding this two-layer network model in which we hereby 
name a few. 

\begin{itemize}
\item \textbf{Cyber-physical social networks.}  If edges are interpreted as the friendship between persons, the corresponding nodes bond with each other in online social networks, physical networks, or both, thus forming a multi-layer social network. 
\item \textbf{Social-hub.} A user employs the same account for two different social networks so that many such users link them into one giant two-layer social network. 

\end{itemize} 

In the following, for the convenience of narration, we regard two nodes connected by an edge in $ \mathcal{E}_C$ as identical. That means if $(u_1,u_2)$ is in $ \mathcal{E}_C$, we take $u_1$ and $u_2$ as one node $u$. 
For each edge $(u, v)$, we define the \emph{neighbors of an edge} (or \emph{neighboring edges} interchangeably) as the extra edges connecting either node $u$ or $v$ but not both in the whole network.  Let $b_u$ be the blue degree of node $u$ which means the number of blue nodes adjacent to u, and let 
$r_u$ be the corresponding red degree of u. For the blue edge $(u, v)$, we denote by $b(u,v)$ the number of its 
blue neighbors and by $r(u, v)$ the number of its red neighbors.

There are
\begin{eqnarray}
b(u,v) &:=& b_u + b_v - 2\mathbb{I}_{E_B(u,v)}, \nonumber\\
r(u, v) &:=& r_u + r_v - 2\mathbb{I}_{E_R(u,v)}. \nonumber
\end{eqnarray}

where $\mathbb{I}_{E_R(u,v)}$ is 1 if the red edge $E_R(u,v)$ exists and 0 otherwise. 

Note that the parameters $b(u,v)$ and $r(u,v)$ will be used to calculate state transition probabilities in the random walk.

\subsection{Two-layer Graphlets}
Graphlets are defined as small induced subgraphs of a large network, where an induced subgraph means 
that once some nodes are selected, all the edges between them are selected too. 
Knowing graphlet statistics is of great importance in network science and engineering. Extensive efforts have been 
devoted to the understanding on how social relationship patterns are formed and evolve in online social networks \cite{huang2015triadic,dong2012link,klimek2013triadic}. However, an important yet largely overlooked problem is to measure graphlet concentration in 
multi-layer networks. The graphlets in the above two-layer networks simultaneously capture the interaction between users in 
both the online social network and the physical world or another social network.

An induced graph of $\mathcal{G}$, $\mathcal{G}' = (\mathcal{V}', \mathcal{E}')$, is a connected 
subgraph whose vertices and edges are all in $\mathcal{G}$, i.e. $\mathcal{E}' =\{(u,v):u,v \in \mathcal{V}', (u,v) \in \mathcal{E}\}$. Let us define $C^{(k)}$ as the set of all connected and induced subgraphs (CISes) containing $k$ nodes. 
Consider two graphs $\mathcal{G}'_1$ and $\mathcal{G}'_2$. If there exists a bijection $\varphi:V'_1 \rightarrow 
V'_2$ with $(u,v)\in  \mathcal{E}'_1 \Leftrightarrow (\varphi(u), \varphi(v))\in  \mathcal{E}'_2$, we claim that 
$\mathcal{G}'_1$ and $\mathcal{G}'_2$ are isomorphic. The two isomorphic graphlets are deemed as the same type of 
graphlet. 

\begin{figure}[h]
\centering
\includegraphics[width = .8\textwidth]{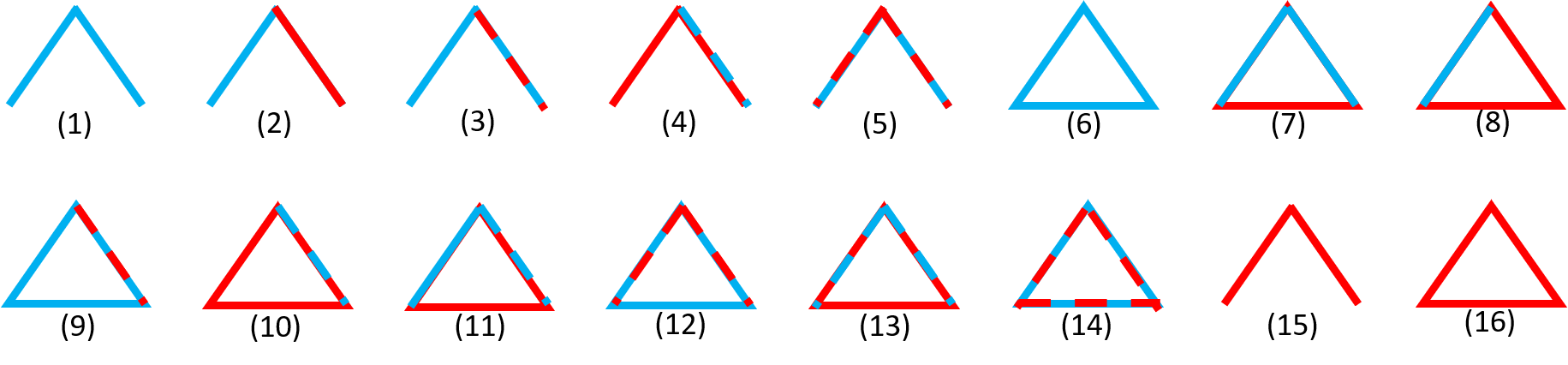}
\caption{\textbf{An illustration of the two-layer graphlets}}
\label{fig:two_layer_graphlets}
\end{figure}

After grouping isomorphic CISes, $C^{(k)}$, is partitioned into $N_k$ classes in which $C_i^{k}$ refers to the $i^{th}$ type of 
isomorphic CISes. Different $k$ yields different $N_k$ that grows exponentially with regard to $k$. 
The number of non-isomorphic classes is two when $k$ is three, and it grows to six when $k$ is four. The non-isomorphic graphlets in our two-layer graph $\mathcal{G}$ is more complicated, embracing richer 
representations among local nodes. Fig. \ref{fig:two_layer_graphlets} illustrates all sixteen graphlets with different edge colors. 
Each graphlet consists of \emph{three} nodes in two layers where two nodes with an inter-layer link are deemed as the same node. An edge colored in \textbf{BLUE+RED} means that the two nodes are connected at the both layers. 
Note that $C_1^{(3)}$ and $C_6^{(3)}$ contain only blue edges, and $C_{15}^{(3)}$ and $C_{16}^{(3)}$ contain only 
red edges.  In this paper, we focus on the case $k=3$ so that 
the superscript $k$ is ignored, but our analytical framework accommodates the cases $k>3$. The concentration of graphlets is defined as
\begin{eqnarray}
d_i = \frac{\big| C_{i} \big|}{\big| {C} \big|}
\end{eqnarray}
where $\big| C_{i} \big|$ is the number of type $i$ CISes and $\big|{C}\big|$ is the total number of CISes. 

\subsection{Two-layer Graphlet Sampling}

New challenges arise in the sampling of a multi-layer network besides the unavailability of complete network topology. \emph{Not all layers can be sampled in the same way.} For instance, an OSN layer can be queried 
through a random walk approach. After querying a node or an edge, this approach is able to jump to one of its 
neighbors. Such a random walk on physical person-to-person networks or privacy sensitive social networks is either infeasible or costly. For instance, a 
crawler can visit a person's direct connections, but not those beyond one-hop. 
Furthermore, querying the friendship relationships in physical 
world has to pay some coupons or needs multiple rounds of interactions. 
We consider a novel sampling problem in restricted and multi-layer networks, 
in which the first layer allows random walk, while the second layer only supports node sampling from the nodes already visited 
at the first layer. 

\noindent\textbf{Random Walk Sampling:} A random walk sampling over graph $\mathcal{G}$ is a process that enables the moving 
from a node or an edge to one of its neighbors chosen uniformly at random, and that starts from an initial node and terminates 
until certain stopping criteria.

\noindent\textbf{Node/Edge Sampling:} A target fraction of nodes/edges are chosen independently and uniformly at
random for inclusion in graph $\mathcal{G}$, and the attached edges/nodes to these nodes/edges 
are included to construct the \emph{induced subgraph}.

Our graph sampling problem differs from the literature in two aspects. Firstly, the random walk is feasible only at one layer, and the node sampling depends on the result of random walk. Secondly, our random walk procedure is mixed up with sampling to obtain some information from the restricted layer. The major notations are summarized in Table \ref{notations}.

\begin{table}[htpb]
\centering
\caption{Notations}\label{notations}
\begin{tabular}{cl}
\toprule
$\mathcal{G}$&a two-layer complex network\\
\hline
$\mathcal{V}$&set of all nodes, including nodes in blue and red levels\\
\hline
$\mathcal{E}$&set of all edges, including blue and red edges\\
\hline
$\mathcal{V}_B$&set of nodes in blue level\\
\hline
$\mathcal{E}_B$&set of blue edges\\
\hline
$\mathcal{E}_C$&set of edges connecting two levels\\
\hline
$|C_i|$&exact number of $i_{th}$ graphlet\\
\hline
$d_i$&exact concentration of $i_{th}$ graphlet\\
\hline
$\hat{|{C}_i|}$&estimator of $|C_i|$\\
\hline
$\hat d_i$&estimator of $d_i$\\
\hline
$X_m$&one blue edge or one blue node\\
\hline
$Y_{m+1}$&one red edge or one red node, specifically, it is adjacent to $X_m$\\
\hline
$b_{X_m}$&number of blue edges/nodes which are adjacent to $X_m$\\
\hline
$r_{X_m}$&number of red edges/nodes which are adjacent to $X_m$\\
\hline
$W$&state space of our Markov chain\\
\hline
$\pi$&stationary distribution of our Markov chain\\
\hline
$P$&state transition matrix of our Markov chain\\
\hline
$\tau(\epsilon)$&mixing time of our Markov chain\\
\hline
$\alpha_i$&number of states which are corresponded to $i_{th}$ graphlet\\
\bottomrule
\end{tabular}
\end{table}

\section{Design of Sampling Algorithms}\label{sec:algorithm}
\label{sec:design}

In this section, we present a random walk framework for graphlet estimation in restricted two-layer networks (a two-layer network as a typical instance of multi-layer network).
Both the node-based and the edge-based random walks are investigated.

\subsection{Node-by-node Random Walk}

Recall that a graphlet contains three nodes in one or two layers. We induce the graphlets by packing three 
nodes as a 3-tuple. The restricted node-by-node random walk (\textsf{RWNbN}) operates as the following. 

\begin{figure}[h] 
\centering
\includegraphics[width = 4in]{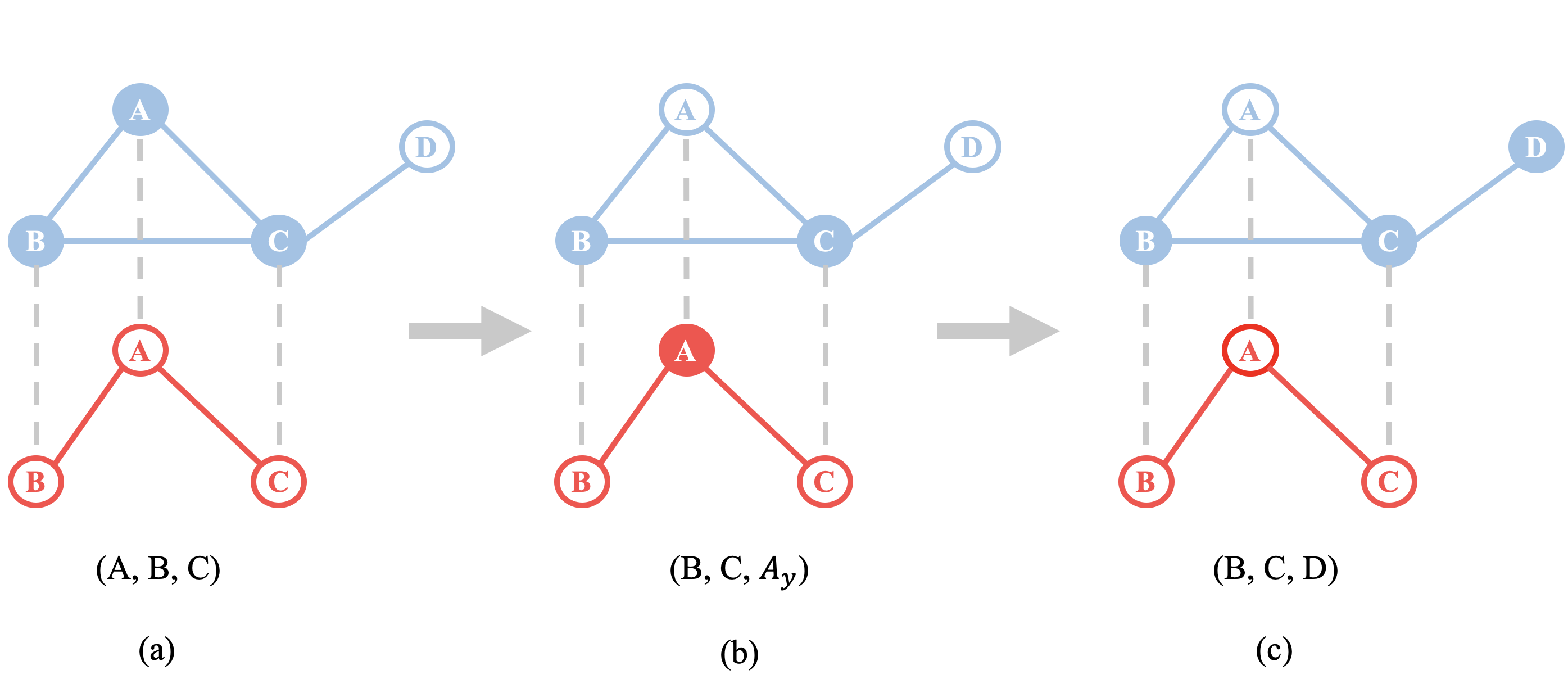}
\caption{\textbf{An illustration of node-by-node random walk}}
\label{fig:temp_transition_nrw}
\end{figure}

The initial 3-tuple is $(A,B,C)$ at Fig.\ref{fig:temp_transition_nrw} (a) with all the nodes in blue. In the next step, the random walker picks a red neighbor of $C$, that is $A$, with probability $\frac{1}{4}$ (because  $b_C = 3$ and $r_C = 1$). 
This makes the random walker enter the red layer as Fig.\ref{fig:temp_transition_nrw} (b). We denote $A_y$ as the node $A$ at the 
red layer. The 3-tuple now turns into $(B,C,A_y)$. At the third step, owing to the restriction of accessing $A_y$'s neighbors, the random walker must return 
to the blue layer, yielding a new tuple $(B,C,D)$ shown in Fig.\ref{fig:temp_transition_nrw} (c) with probability $\frac{1}{3}$.

Formally, a node $u$ is visited at the current step, the node-by-node random walk (\textsf{RWNbN}) drives the sampler to 
visit the node $v$ at the next step where $v$ is a neighbor of $u$. 
Let $X_m$ be the $m^{th}$ node at the blue layer and let $Y_{m+1}$ be the randomly sampled neighbor of $X_m$
at the red layer if it exists. 
Define a 3-tuple $S_t$ as a \emph{\textbf{state}} of random walk that consists of the three
most recently traversed nodes at the $t^{th}$ step. Note that three visited 
nodes can induce a 3-node graphlet. 
For clarity, there has $S_t:=(X_{m-1}, X_m, X_{m+1})$ if all the nodes are in the blue graph, and $S_t:=(X_{m-1}, X_m, Y_{m+1})$ if 
the third node is at the red layer via node sampling. 
Meanwhile, a state cannot include two red nodes 
because only one-hop neighbor at the red layer is allowed to visit. Define the state space as $W$ = \{$(X_m,X_{m+1},X_{m+2})\cup(X_m,X_{m+1},Y_{m+2}), \;\forall m=1,2,\cdots$\} in our sampling. 

The random walk on a graph constitutes an ergodic Markov process. 
We next derive the state transition probabilities of the restricted random walk.
At a given time step $S_t=(X_{m-1}, X_m, X_{m+1})$, 
we randomly pick a node among $b_{X_{m+1}}$ blue and $r_{X_{m+1}}$ red ones to visit. With probability $\frac{1}{r_{X_{m+1}}+b_{X_{m+1}}}$ the random 
walk moves to a blue node  $X_{m+2}$ and yields state $S_{t+1}=(X_m, X_{m+1},X_{m+2})$ at the next time step, 
and with probability $\frac{1}{r_{X_{m+1}}+b_{X_{m+1}}}$ it moves to an 
adjacent red node and reaches the state $S_{t+1}:=(X_m, X_{m+1}, Y_{m+2})$. At a given state $S_t = (X_m,X_{m+1},Y_{m+2})$, because the red graph does not allow random walk, 
a blue node adjacent to $X_{m+1}$, namely $X_{m+2}$, is chosen uniformly so that the state at the $(t+1)^{th}$ step is expressed as 
$S_{t+1}:=(X_m, X_{m+1},X_{m+2})$ with probability $\frac{1}{b_{X_{m+1}}}$. We should notice that $Y_{m+2}$ is not necessarily the corresponding node of $X_{m+2}$ in the red layer. We summarize the state transition probabilities in Table \ref{table:state_trans_node}.

\begin{table}[htpb]
\caption{State transition matrix of node-by-node RW}\label{table:state_trans_node} 
\centering
\begin{tabular}{|c|c|c|}
\hline
\diagbox{Current}{Next}&$(X_m,X_{m+1},X_{m+2})$&$(X_m,X_{m+1},Y_{m+2})$\\ 
\hline
$(X_{m-1},X_m,X_{m+1})$&$\frac{1}{r_{X_{m+1}}+b_{X_{m+1}}}$&$\frac{1}{r_{X_{m+1}}+b_{X_{m+1}}}$\\
\hline
$(X_{m},X_{m+1},Y_{m+2})$&$\frac{1}{b_{X_{m+1}}}$&0\\
\hline
\end{tabular}
\end{table}

\subsubsection{Unbiased Estimator} 

The random walk on graph $\mathcal{G}$ does not automatically 
reveal the graphlet concentration. We need to derive the stationary distribution of our Markov chain, and 
map Markovian states into graphlets. 

\noindent\textbf{Stationary Distribution.} 
Our restricted random walk process in a two-layer network is an irreducible Markov chain. For any two states 
$(X_m,X_{m+1},X_{m+2})$ and $(X_n,X_{n+1},X_{n+2})$, they can reach each other on the blue graph; for any two states 
$(X_m, X_{m+1},Y_{m+2})$ and $(X_n, X_{n+1},Y_{n+2})$, they can reach each other through the edges of the blue graph. So can states $(X_m,X_{m+1},X_{m+2})$ and $(X_n,X_{n+1},Y_{n+2})$.
It is well-known that any finite and irreducible Markov chain has a unique stationary distribution on all the states. 

We next compute the stationary distribution $\mathbf{\pi}$ as the following
\begin{eqnarray}
\begin{cases}
\pi(X_m,X_{m+1},X_{m+2}) = \frac{1}{Mb_{X_{m+1}}}\\
\pi(X_m,X_{m+1},Y_{m+2}) = \frac{1}{M(r_{X_{m+1}}+b_{X_{m+1}})}
\end{cases}
\end{eqnarray}
where $M = 2|\mathcal{E}_B| + \sum_{v \in \mathcal{V}_B}\frac{b_v r_v}{b_v + r_v}$. $\mathcal{E}_B$ is the edge set of blue layer, and $\mathcal{V}_B$ is the node set of blue layer. The detailed analysis can be found in Appendix A. One can observe that $\pi(X_m,X_{m+1},X_{m+2})$ is larger than $\pi(X_m,X_{m+1},Y_{m+2})$ because the sampling of a red edge is possible 
only when the corresponding blue edge has been sampled by random walk.

\noindent\textbf{Isomorphic State Precomputation.} 
We hereby provide a mapping method from traversed states to 3-node graphlets. A key observation is that a graphlet 
corresponds to several Markovian states in the random walk. Hence, before estimating the graphlet concentration, we 
need to figure out the relationship between states and graphlets. 

Note that  any two states forming the same graphlet are called \emph{\textbf{isomorphic states}}. 
The isomorphic state coefficient refers to the number of isomorphic states in correspondence with the same graphlet.  
The isomorphic state coefficient can be computed in advance. Its precomputation in a multi-layer network, different from that in 
a single layer one, depends on both the property of the graphlet and the restricted random walk algorithm. We use two examples to highlight their differences.

\noindent\textbf{Example 1.} If we perform a restricted random walk on the sixth graphlet consisting of three blue edges, and denote three nodes by $u,v$ and $w$.
Different traverse trajectories result in six states respectively: $(u,v,w)$, $(u,w,v)$, $(v,u,w)$, $(v,w,u)$, $(w,u,v)$ and $(w,v,u)$. The isomorphic state coefficient is 6 accordingly.

\noindent\textbf{Example 2.} If a restricted random walk happens on the tenth graphlet, there are only two isomorphic states: 
$(u,v,w)$ and $(v,u,w)$, where two end nodes of blue edge are denoted by $u,v$ and the third node is $w$ . Since only the node sampling on the red graph is allowed, two blue nodes should be visited first in our restricted random walk on the blue graph, leading to the isomorphic state coefficient of two. 

We use $\alpha_i$ to denote the isomorphic state coefficient of the $i^{th}$ graphlet, 
and summarize all the coefficients of 3-node graphlets in Table \ref{table:isomorphic_state}. Intuitively, a larger coefficient 
means more blue edges in the blue graph that supports random walk. 

\begin{table}[htb]
\centering
\label{tab5}
\begin{tabular}{|c|c|c|c|c|c|c|}
\hline
$\alpha_1$&$\alpha_2$&$\alpha_3$&$\alpha_4$&$\alpha_5$&$\alpha_6$&$\alpha_7$\\ 
\hline
2&1&3&1&4&6&4\\
\hline
$\alpha_8$&$\alpha_9$&$\alpha_{10}$&$\alpha_{11}$&$\alpha_{12}$&$\alpha_{13}$&$\alpha_{14}$\\
\hline
2&8&2&5&10&6&12\\
\hline
\end{tabular}
\caption{Isomorphic state coefficient $\alpha_i$}
\label{table:isomorphic_state}
\end{table}

\noindent\textbf{Estimator.} 
Denote by $g_i$ an indicator function of an arbitrary 
Markovian state $S$ that has
\begin{eqnarray}
g_i (S) = \begin{cases}
1 \quad\quad \text{$S$ \ induces the $i^{th}$ graphlet}\\
0 \quad\quad \text{otherwise}
\end{cases}.
\end{eqnarray}
Summing all the possible states and mapping 
them into different graphlets, there exists 
$\sum_{S\in W}g_i(S) = \alpha_i |C_i|$.
The parameter $\alpha_i$ means that a graphlet has 
multiple isomorphic copies in the state space. 
The random walk operations
harvest $n$ states, $\{S_j\}_{j=1}^N\subset W$.

The unbiased estimation is put on the foundation of 
Strong Law of Large Numbers (SLLN) \cite{Geyer2005MarkovCM}. 
If an irreducible Markov chain has finite state space $W$ with a stationary distribution $\mathbf{\pi}$, and there exists a function of the state $S$, $h(S):W\rightarrow \mathbb{R}$, the expectation of $h(S)$ at all the states can be defined as 
\begin{eqnarray}
\mu = \mathbb{E}_{\pi}[h(S)] = \sum_{S \in W}h(S)\pi(S).
\end{eqnarray}
Given $n$ states $\{S_j\}_{j=1}^n$, SLLN provides the 
following theorem.
\begin{theorem}
\label{theorem:unbiased}\cite{Geyer2005MarkovCM}
$\frac{1}{n}\sum_{j=1}^n h(S_j) \stackrel{a.s.}{\longrightarrow}\mathbb{E}_{\pi}[h]$ as $n \to \infty$.
\end{theorem}
We can obtain
\[\begin{aligned}
\frac{1}{n}\sum_{j=1}^n\frac{g_i(S_j)}{\pi(S_j)}\stackrel{a.s.}{\longrightarrow}\mathbb{E_{\pi}}\left[\frac{g_i}{\pi}\right] = \sum_{S \in W}\frac{g_i(S)}{\pi(S)} \times \pi(S),
\end{aligned}\]
and the last term is actually $\sum_{S  \in  W}g_i(S) = \alpha_i |C_i|$.
Thus, $\hat {|C_i|} = \frac{1}{n} \sum_{j=1}^n \frac{g_i(S_j)}{\alpha_i\pi(S_j)}$ is an unbiased estimator of the number of the $i^{th}$ graphlet. Our goal is to compute the concentration of the $i^{th}$ graphlet, denoted by $d_i$, $d_i = \frac{|C_i|}{\sum_{j=1}^{14}|C_j|}$. Because $\hat{|C_i|}$ estimates $|C_i|$ without bias, the unbiased estimator of the concentration of the $i^{th}$ graphlet is given by:
\begin{equation}
\hat d_i = \frac{\hat {|C_i|}}{\sum_{j=1}^{14}\hat {|C_j|}} = \frac{\sum_{j=1}^n\frac{g_i(S_j)}{\alpha_i \pi(S_j)}}{\sum_{j=1}^n\sum_{i=1}^{14}\frac{g_i(S_j)}{\alpha_i \pi(S_j)}}.
\end{equation}

Note that the stationary distribution $\pi(S_j)$ contains an unknown variable $M$. 
This variable is pertinent to the global information of the two-layer graph, and 
can not be known as a priori. Fortunately, $M$ appears at both the numerator and denominator 
so that it is canceled out at the computation of $\hat d_i$. 

Algorithm \ref{RWNbN} specifies the procedure of obtaining an unbiased estimator. When we calculate $\pi(S_j)$, $M$ is set aside actually. 

\begin{algorithm}
\caption{Node-by-node Random Walk: \textsf{RWNbN}} 
\label{RWNbN}
\begin{algorithmic}
\REQUIRE two-layer network: $\mathcal{G}$, number of samples: $n$
\ENSURE estimated graphlet concentration $\hat d_i$
\STATE $\hat{|C_i|}$ = 0
\STATE Randomly pick a valid initial state $S_1 = (X_1,X_2,X_3)$ 
// $X_1$ and $X_2$ and $X_3$ are nodes in blue level which induce a 3-node graphlet
\STATE Random walk counter t = 1
\WHILE{t $\leq$ n} 
\STATE $i$ = type of the subgraph induced by $S_t$
\STATE $\hat {|C_i|} = \hat {|C_i|} + \frac{1}{\alpha_i \pi(S_t)}$
\STATE $t = t + 1$
\IF{$S_t$ is in the form $(X_m, X_{m+1}, X_{m+2})$} 
\STATE Uniformly pick a node in the blue layer adjacent to $X_{m+2}$ with probability $\frac{1}{r_{X_{m+2}}+b_{X_{m+2}}}$, 
or a node in the red layer adjacent to node $X_{m+2}$ with probability $\frac{1}{r_{X_{m+2}}+b_{X_{m+2}}}$
\ELSE

\STATE Uniformly pick a node in the blue layer which is a neighbor of $X_{m+1}$
\ENDIF 
\ENDWHILE
\STATE $\hat d_i = \frac{\hat {|C_i|}}{\sum_{j=1}^{14}\hat {|C_i|}}$, $i \in \{1,2,...,14\}$\\
\end{algorithmic}
\end{algorithm}.

\subsubsection{Error Bound}
We next analyze the error bound of our random walk algorithm, i.e. how many steps of random 
walk are required in order to guarantee 
a given estimation accuracy. 

\begin{theorem}
\label{theorem:errorbound}\cite{Chung2012ChernoffHoeffdingBF}
Let MC be an ergodic Markov chain with state space $W$ and stationary distribution $\pi$. Let $\tau = \tau(\zeta)$ be its $\zeta-mixing$ time for $\zeta \leq \frac{1}{8}$. Let $\{S_j\}_{j=1}^n$ denote a n-step random walk on MC starting from an initial distribution $\phi$ on $W$, i.e., $S_1 \leftarrow \phi$. For every $i \in \{1,2,...,n\}$, let $f_i : W \rightarrow [0,1]$ be a weight function at step $i$ such that the expected weight $\mathbb{E}_\pi[f_i] = \mu$ for all i. Define the total weight of the walk $\{S_j\}_{j=1}^n$ by $Z = \sum_{i=1}^n f_i(S_i)$. There exists a constant $c$ (which is independent of $\mu$, $\epsilon$ and $\zeta$) such that
\begin{equation}
Pr[| \frac{Z}{n}-\mu |> \epsilon] \leq \ c||\phi||_{\pi}exp(-\epsilon^2\mu n/ (72\tau))
 \end{equation}
 for $0 < \epsilon < 1$.
 \end{theorem}

Define $H = \max_{S \in W}\frac{1}{\pi(S)}$ and $\alpha_{min}$ = $\min_{1\leq i \leq14}\alpha_i$. Denote by $\tau(\zeta)$ the mixing time of our Markov chain and denote by $\phi$ the initial distribution of the visited states with 
\begin{equation*}
||\phi||_{\pi} = \sum_{S \in W}\frac{\phi^2(S)}{\pi(S)}.
\end{equation*}
Let $\Lambda_i = min\{\alpha_i |C_i|, \alpha_{min} |C|\}$, where $|C| = \sum_{i = 1}^{14}|C_i|$. The subscript $i$ is dropped 
when we do not specify the type of the graphlet. 
The following theorem guarantees that the relative 
error is below a sufficiently small $\epsilon$ as the 
number of random walk steps is greater than a
certain threshold. 

\begin{theorem}
\label{ErrorBound}
$\forall$ $0 < \delta < 1, \exists$ constant $\xi$, such that, when $n \geq \xi \frac{H}{\Lambda} \frac{\tau}{\epsilon^2}\ln \frac{||\phi||_{\pi}}{\delta}$, we have 
\begin{equation}
Pr((1-\epsilon)d_i \leq \hat d_i \leq (1+\epsilon)d_i) > 1-\delta.
\end{equation}
\end{theorem}
The detailed proofs can be found in Appendix B.

There are two parameters $H$ and $\Lambda$ in the inequality with regard to the
threshold $n$ that influence the convergence rate 
of a sampling algorithm. Note that $H$ is determined 
by the specific sampling method, and $\Lambda$ depends on the graphlet count of a dataset. 
Thus, the sampling of a dense graphlet needs fewer random walk steps to converge, while that of a rare one demands 
many more steps.

\subsection{Edge-by-edge Random Walk}

Similarly, we can induce 3-node graphlets by packing two sampled edges as a two-tuple. 
Suppose that an edge $(u,v)$ is visited at the current step, the edge-by-edge random walk drives the sampler to 
visit the edge $(u', v')$ at the next step where either $u'$ or $v'$ remains unaltered, i.e. $u=u'$ or $v=v'$, but not both. 
For convenience, we use the same notation as the previous algorithm. We let $X_m$ be the $m^{th}$ edge of the blue graph and let $Y_{m+1}$ be the neighbor of the corresponding edge of 
the red graph if it exists. With certain abuse of notations, we define a two-tuple $S_t$ as a \emph{\textbf{state}} of random walk that consists of the two 
most recently traversed edges. There is $S_t:=(X_{m}, X_{m+1})$ if both edges are in blue, and $S_t:=(X_m, Y_{m+1})$ if 
a red edge is just visited via edge sampling. Our design of state resounds to two important properties. One is that 
the two edges are adjacent to each other, the other is that an edge at the red graph can only be reached through its 
neighbour at the blue graph, and must return to the blue graph afterwards. 
Let the state space be $W = \{(X_m,X_{m+1})| X_m \cap X_{m+1}\neq \varnothing , \forall m\} \cup \{(X_m,Y_{m+1}) | X_m \cap Y_{m+1} \neq \varnothing, \forall m\}$.

\begin{figure}[!ht]
\centering
\includegraphics[width = 4in]{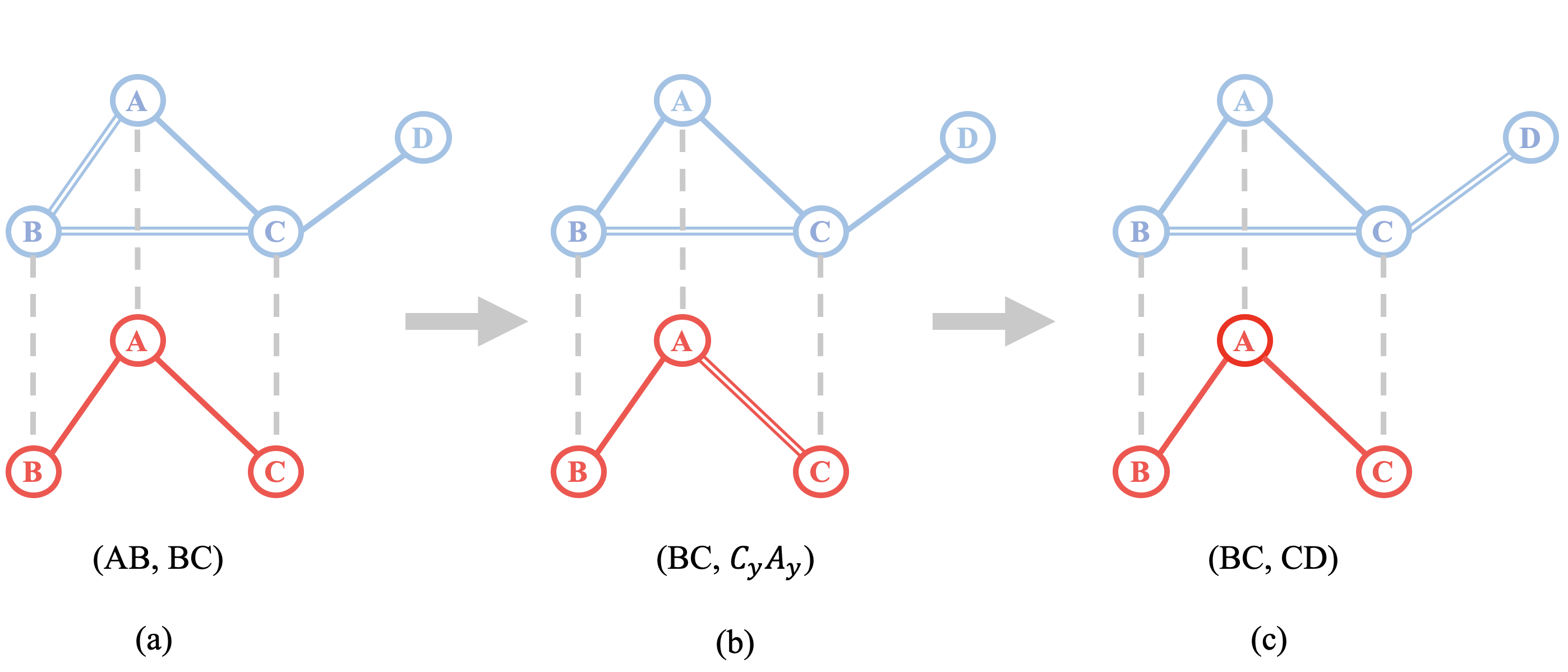}
\caption{\textbf{An illustration of edge-by-edge random walk}}
\label{fig:temp_transition_erw}
\end{figure}

To better understand the operations of edge-by-edge random work, we illustrate the 
procedure in Fig.\ref{fig:temp_transition_erw}. A hollow line denotes an edge at the 
current state, e.g. $(AB, BC)$ in Fig.\ref{fig:temp_transition_erw} (a). If the random walker 
enters the red layer via node $C$ and samples the red edge $C_yA_y$, the new state turns into 
$(BC, C_yA_y)$ in Fig.\ref{fig:temp_transition_erw} (b). Due to the restriction at the red graph, the random walker returns to the 
blue layer (i.e. to node $C$) and visits a new blue edge $CD$ that yields a new state 
$(BC, CD)$ in Fig.\ref{fig:temp_transition_erw} (c).

We next derive the state transition probabilities of the restricted random walk. At a given state $S_t:=(X_{m}, X_{m+1})$, 
we randomly pick an edge among $b_{X_{m+1}}$ blue and $r_{X_{m+1}}$ red ones to visit. With probability $\frac{1}{r_{X_{m+1}}+b_{X_{m+1}}}$ the random 
walk moves to a blue edge $X_{m+2}$ and yields state $S_{t+1}:=(X_{m+1}, X_{m+2})$, 
and with probability $\frac{1}{r_{X_{m+1}}+b_{X_{m+1}}}$ it moves to an 
adjacent red edge and yields state $S_{t+1}:=(X_m, Y_{m+1})$. If the current state is $S_{t} = (X_m, Y_{m+1})$, because the red graph does not allow random walk, 
a blue edge adjacent to $X_m$, namely $X_{m+1}$, is chosen uniformly so that the new state is expressed as 
$S_{t+1}:=(X_m, X_{m+1})$ with probability $\frac{1}{b_{X_{m}}}$. 
For clarity, we summarize the state transition probabilities in Table \ref{table:state_trans_edge}.
\begin{table}[htpb]
\centering
\begin{tabular}{|c|c|c|}
\hline
\diagbox{Current state}{Next state}&$(X_{m+1},X_{m+2})$&$(X_{m+1},Y_{m+2})$\\ 
\hline
$(X_{m},X_{m+1})$&$\frac{1}{r_{X_{m+1}}+b_{X_{m+1}}}$&$\frac{1}{r_{X_{m+1}}+b_{X_{m+1}}}$\\
\hline
$(X_{m+1},Y_{m+2})$&$\frac{1}{b_{X_{m+1}}}$&0\\
\hline
\end{tabular}
\caption{State transition matrix of edge-by-edge RW}
\label{table:state_trans_edge}
\end{table}

\noindent\textbf{Stationary Distribution.} 
We direcly show the stationary distribution $\mathbf{\pi}$ as the following
\begin{eqnarray}
\begin{cases}
\pi(X_m,X_{m+1}) = \frac{1}{M}\\
\pi(X_m,Y_{m+1}) =  \frac{b_{X_{m}}}{M (r_{X_{m}}+b_{X_{m}})}

\end{cases}
\end{eqnarray}
where $M = 2|\mathcal{E}_B| + \sum_{v \in \mathcal{V}_B}\frac{b_v r_v}{b_v + r_v}$. The detailed analysis is very similar to the proof of node by node random walk in Appendix A. 
One can observe that $\pi(X_m,X_{m+1})$ is larger than $\pi(X_m,Y_{m+1})$ because the sampling of a red edge is possible 
only when the corresponding blue edge has been sampled by random walk. 
Because the mapping from two edges to a 3-node graphlet is the same as that from three nodes to a 3-node graphlet, the precomputations of isomorphic are same. We directly show the procedure of 
edge-by-edge random walk (\textsf{RWEbE}) in Algorithm \ref{RWEbE}.

The connection between \textsf{RWEbE} and \textsf{RWNbN} is the following. For a given graph, we can create a new 
relationship graph in which each new node refers to an edge in the original graph, and two nodes are connected if  
their corresponding edges are adjacent in the original graph. Therefore, \textsf{RWEbE} is actually the node-by-node random 
walk on this relationship graph. The relationship graph is of larger size since there are more edges than nodes in the 
original graph, and the average degree is expected to be larger. In a general single layer graph, it is hard to tell whether \textsf{RWEbE} or 
\textsf{RWNbN} is more accurate on a specific graphlet. When constructing a relationship graph on the 
two-layer restricted network, we magnify the blue layer much more than the red layer since the access to red nodes is limited 
to be within one-hop. Hence, the chance of visiting the graphlets with more blue nodes is higher, and the chance of 
meeting those with more red nodes is lower. The accuracy of graphlet concentration is influenced accordingly. 
The convergence property of \textsf{RWEbE} is guaranteed by 
Theorem \ref{ErrorBound}. The convergence rate largely depends on how rare a graphlet is in a network.

\begin{algorithm}
\caption{Edge-by-edge Random Walk : \textsf{RWEbE}} 
\label{RWEbE}
\begin{algorithmic}
\REQUIRE two-layer network: $\mathcal{G}$,  random walk steps: $n$
\ENSURE estimated graphlet concentration: $\hat d_i$
\STATE $\hat {|C_i|}$ = 0, \; $\forall i$; Random walk counter $t=1$
\STATE Randomly pick an initial state $S_1 = (X_1,X_2)$ \\
//$X_1$ and $X_2$ are blue edges with one common node
\WHILE{$t \leq n$} 
\STATE $i$ = type of the subgraph induced by $S_t$
\STATE $\hat {|C_i|} = \hat {|C_i|} + \frac{1}{\alpha_i \pi(S_t)}$
\STATE $t = t + 1$
\IF{$S_t$ is in the form $(X_m, X_{m+1})$} 
\STATE Uniformly pick a blue edge adjacent to $X_{m+1}$ with probability $\frac{1}{r_{X_{m+1}}+b_{X_{m+1}}}$ or 
a red edge adjacent to $X_{m+1}$ with probability $\frac{1}{r_{X_{m+1}}+b_{X_{m+1}}}$
\ELSE
\STATE Uniformly pick a blue edge adjacent to $X_m$
\ENDIF 
\STATE Update $S_t$
\ENDWHILE
\STATE $\hat d_i = \frac{\hat {|C_i|}}{\sum_{j=1}^{14}\hat {|C_i|}}$, $i \in \{1,2,...,14\}$\\
\end{algorithmic}
\end{algorithm}

\section{Sampling Budget Adjustment}\label{sec:extension}
An interesting property of sampling graphlets in two-layer networks is that ``\emph{all graphlets in different layers are not sampled equally}''. 
For instance, the random walker enters the red layer through the blue layer so that 
the latter may be visited more frequently. Even though our proposed sampling algorithms 
are unbiased with provable error bounds, one can observe that 
the estimation accuracy of the graphlets with more blue edges is better off than that with 
more red edges. Given the same total random walk steps, one can choose to distribute 
more steps to a particular layer so as to improve the sampling accuracy of some graphlets, whereas at the cost of degraded accuracy of the others. This is a new phenomenon that 
has not appeared in traditional single-layer networks. In this section, we consider the 
distribution of sampling budget in two layers and explore the tradeoff on the accuracy of
different graphlets.

\subsection{Sampling More Graphlets with More Red Edges}
Previously, when the sampling is at a state $(X_m,X_{m+1},Y_{m+2})$, it will go back to the blue layer and transit to 
a new state $(X_m,X_{m+1},X_{m+2})$. In the new setting, 
the sampler can visit one more red node, i.e. transiting to the state $(X_{m+1},Y_{m+2},Y_{m+3})$, before returning to the 
blue layer. In other words, when there is an opportunity to enter the red network, the red nodes are sampled twice. 
Our sampling framework applies to this new situation. That means the stationary distribution, state transition matrix, isomorphism coefficients and the corresponding sampling algorithm namely \textsf{RWOMRN} (\textsf{OMRN} means one more red nodes) should be modified, but all the proof of the unbiasedness and error bound remains same.

\noindent\textbf{Stationary Distributions.}
Our first step is to compute the stationary distribution of all the Markovian states.

\[ \begin{cases}
\pi(X_m,X_{m+1},X_{m+2}) = \frac{1}{Mb_{X_{m+1}}}\\
\pi(X_m,X_{m+1},Y_{m+2}) = \frac{1}{M(r_{X_{m+1}}+b_{X{m+1}})}\\
\pi(X_m,Y_{m+1},Y_{m+2}) = \frac{b_{X_m}}{M(r_{X{m}}+b_{X_{m}})r_{Y_{m+1}}}
\end{cases}, \]
where $M =  \sum_{v \in \mathcal{V}_B}\frac{b_v^2-b_v-r_v+3b_v r_v}{b_v + r_v}$,  $\mathcal{V}_B$ is the node set of blue level.

\noindent\textbf{State transition matrix.}

\begin{table}[htpb]
\centering
\begin{tabular}{|c|c|c|c|}
\hline
\diagbox{Current state}{Next state}&$(X_m,X_{m+1},X_{m+2})$&$(X_{m},X_{m+1},Y_{m+2})$&$(X_{m+1},Y_{m+2},Y{m+3})$\\ 
\hline
$(X_{m-1},X_{m},X_{m+1})$&$\frac{1}{r_{X{m+1}}+b_{X_{m+1}}}$&$\frac{1}{r_{X_{m+1}}+b_{X_m+1}}$&0\\
\hline
$(X_{m},X_{m+1},Y_{m+2})$&0&0&$\frac{1}{r_{Y_{m+2}}}$\\
\hline
$(X_{m+1},Y_{m+2},Y_{m+3})$&$\frac{1}{{b_{X_{m+1}}}^2}$&0&0\\
\hline
\end{tabular}
\caption{State transition matrix of  \textsf{RWOMRN}}
\label{table:state_trans_edge2}
\end{table}

The state transition matrix is different from that appears before. Now, when we are in the state $(X_m,X_{m+1},Y_{m+2})$, in next step, we must explore one more red node and obtain a new state with two red nodes and the probability is $\frac{1}{r_{Y_{m+2}}}$. Note that the denominator cannot be zero, because $Y_{m+2}$ has at least one red neighbor which is the corresponding node of $X_{m+1}$ on red layer.
When we are in a state $(X_{m+1},Y_{m+2},Y_{m+3})$, in next step we must return to blue layer by the mechanism of the algorithm. Consequently, two neighbors of $X_{m+1}$ are picked randomly, and we attain a state with three blue nodes.

\noindent\textbf{Isomorphism Coefficients.}

\begin{table}[htb]
\centering
\begin{tabular}{|c|c|c|c|c|c|c|}
\hline
$\alpha_1$&$\alpha_2$&$\alpha_3$&$\alpha_4$&$\alpha_5$&$\alpha_6$&$\alpha_7$\\ 
\hline
2&1&3&3&6&6&4\\
\hline
$\alpha_8$&$\alpha_9$&$\alpha_{10}$&$\alpha_{11}$&$\alpha_{12}$&$\alpha_{13}$&$\alpha_{14}$\\
\hline
4&8&8&7&12&12&18\\
\hline
\end{tabular}
\caption{Coefficient $\alpha_i$}
\end{table}

The result is different from the coefficient for \textsf{RWNbN}. For example, for fifth graphlet, assume those three nodes on blue level are $u, v, w$ from left side to right side and their corresponding nodes in red level are $u', v', w'$. Under \textsf{RWNbN}, there are four ways to obtain this graphlet, $(u,v,w), (w,v,u),(u,v,w'), (w,v,u')$. But under \textsf{RWOMRN}, we have two more ways $(u,v',w'), (w,v',u')$ because we change the random walk mechanism and allow the random walker to sample twice on red level. 

\begin{algorithm}
\caption{Sampling One More Red Node : \textsf{RWOMRN}} 
\label{RWOMRN}
\begin{algorithmic}
\REQUIRE two-layer network $\mathcal{G}$,  random walk steps $n$
\ENSURE graphlets concentration estimation $\hat d_i$
\STATE $\hat {|C_i|}$ = 0
\STATE Randomly pick a valid initial state $S_1 = (X_1,X_2,X_3)$ 
// $X1$ and $X2$ and $X3$ are blue nodes which induce a 3-node graphlet
\STATE Random walk counter t = 1
\WHILE{t $\leq$ n} 
\STATE i = type ID of induced subgraph of $S_t$
\STATE $\hat {|C_i|} = \hat {|C_i|} + \frac{1}{\alpha_i \pi(S_t)}$
\STATE t = t + 1
\IF{$S_t$ is in the form $(X_m, X_{m+1}, X_{m+2})$} 
\STATE Uniformly pick a blue node which is adjacent to $X_{m+2}$ by chance $\frac{1}{r_{X{m+2}}+b_{X{m+2}}}$, then obtain the next state
\STATE Or uniformly pick a red node which is adjacent to $X_{m+2}$ by chance $\frac{1}{r_{X{m+2}}+b_{X{m+2}}}$, then obtain the next state
\ELSIF{$S_t$ is in the form $(X_m, X_{m+1}, Y_{m+2})$}
\STATE Uniformly pick a red node which is adjacent to $Y_{m+2}$ in red level, then obtain the next state
\ELSE
\STATE ($S_t$ is in the form $(X_m, Y_{m+1}, Y_{m+2})$)
\STATE Uniformly pick 2 blue nodes $X_{m-1}$ and $X_{m+1}$ which are adjacent to $X_m$, and obtain the next state
\ENDIF 
\ENDWHILE
\STATE $\hat d_i = \frac{\hat {|C_i|}}{\sum_{j=1}^{14}\hat {|C_i|}}$, $i \in \{1,2,...,14\}$\\
\end{algorithmic}
\end{algorithm}.

\subsection{Mixed Algorithm}
Previously, we have already shown the algorithm of sampling one node in red level and sampling two nodes in red level. The latter has better performance on those graphlets which have more red edges, but the former is better on graphlets with more blue edges. Actually, given a total of $n$ sampling steps, if more red nodes are sampled, the graphlets with more red edges can be estimated more accurately. However, the estimation accuracy of the
graphlets with more blue edges degrades. 
Consequently, we obtain a balance between those two algorithms. Because we want to estimate concentration, so roughly speaking, making the error some kind of evenly can obtain better estimation of concentration.

So, when in the state $(X_m,X_{m+1},Y_{m+2})$, instead of directly sampling a red node which is neighbor of $Y_{m+2}$, we choose to do so by probability, and we can also return to blue level by probability. Consequently, a balance between two former algorithms is achieved. Compared with two former algorithms, we sample moderate number of red nodes as well as blue nodes given a fixed number of sampling steps.

We show the algorithm in algorithm \ref{RWMixed}.

\noindent\textbf{Stationary distribution and Coefficient.}
Similarly, we directly show the stationary distributions of our state space and coefficients.

\[ \begin{cases}
\pi(X_m,X_{m+1},X_{m+2}) = \frac{1}{Mb_{X_{m+1}}}\\
\pi(X_m,X_{m+1},Y_{m+2}) = \frac{1}{M(r_{X_{m+1}}+b_{X_{m+1}})}\\
\pi(X_m,Y_{m+1},Y_{m+2}) = \frac{b_{X_m}}{M(r_{X_m}+b_{X_m})(r_{Y_{m+1}}+b_{X_m})}
\end{cases}, \]
where
\begin{equation} 
M = 2|\mathcal{E}_B| + \sum_{v \in \mathcal{V}_B}\frac{b_v r_v}{b_v + r_v} + \sum_{v \in \mathcal{V}_B}\sum_{u \in RN(v)}\frac{b_v r_u}{(b_v + r_v)(r_u+b_v)}
\end{equation}
$RN(v)$ denotes the set of red neighbors of v.

\begin{table}[htb]
\centering
\begin{tabular}{|c|c|c|c|c|c|c|}
\hline
$\alpha_1$&$\alpha_2$&$\alpha_3$&$\alpha_4$&$\alpha_5$&$\alpha_6$&$\alpha_7$\\ 
\hline
2&1&3&3&6&6&4\\
\hline
$\alpha_8$&$\alpha_9$&$\alpha_{10}$&$\alpha_{11}$&$\alpha_{12}$&$\alpha_{13}$&$\alpha_{14}$\\
\hline
4&8&8&7&12&12&18\\
\hline
\end{tabular}
\caption{Coefficient $\alpha_i$}
\end{table}

\noindent\textbf{State transition matrix}

\begin{table}[!h]
\centering
\begin{tabular}{|c|c|c|c|}
\hline
\diagbox{Current state}{Next state}&$(X_m,X_{m+1},X_{m+2})$&$(X_{m},X_{m+1},Y_{m+2})$&$(X_{m+1},Y_{m+2},Y{m+3})$\\ 
\hline
$(X_{m-1},X_{m},X_{m+1})$&$\frac{1}{r_{X{m+1}}+b_{X_{m+1}}}$&$\frac{1}{r_{X_{m+1}}+b_{X_m+1}}$&0\\
\hline
$(X_{m},X_{m+1},Y_{m+2})$&$\frac{1}{b_{X_{m+1}}+r_{Y_{m+2}}}$&0&$\frac{1}{r_{Y_{m+2}}+b_{X_{m+1}}}$\\
\hline
$(X_{m+1},Y_{m+2},Y_{m+3})$&$\frac{1}{{b_{X_{m+1}}}^2}$&0&0\\
\hline
\end{tabular}
\caption{State transition matrix of  \textsf{RWMix}}
\end{table}

Here, different from state transition matrix of \textsf{RWOMRN}, when in a state $(X_m,X_{m+1},Y_{m+2})$, it is possible to explore one more red node, but we may also directly return to blue layer. The possible candidates are all blue neighbors of $X_{m+1}$ and all red neighbors of $Y_{m+2}$, so the denominator is $b_{X_{m+1}} + r_{Y_{m+2}}$.

\begin{algorithm}
\caption{Mixed Algorithm : \textsf{RWMix}}
\label{RWMixed}
\begin{algorithmic}
\REQUIRE two-layer network $\mathcal{G}$,  random walk steps $n$s
\ENSURE graphlets concentration estimation $\hat d_i$
\STATE $\hat {|C_i|}$ = 0
\STATE Randomly pick a valid initial state $S_1 = (X_1,X_2,X_3)$ 
// $X_1$ and $X_2$ and $X_3$ are blue nodes which  induce a 3-node graphlet
\STATE Random walk counter t = 1
\WHILE{t $\leq$ n} 
\STATE i = type ID of induced subgraph of $S_t$
\STATE $\hat {|C_i|} = \hat {|C_i|} + \frac{1}{\alpha_i \pi(S_t)}$
\STATE t = t + 1
\IF{$S_t$ is in the form $(X_m, X_{m+1}, X_{m+2})$} 
\STATE Uniformly pick a blue node which is adjacent to $X_{m+2}$ by chance $\frac{1}{r_{X_{m+2}}+b_{X_{m+2}}}$, then obtain the next state
\STATE Or uniformly pick a red node which is adjacent to $X_{m+2}$ by chance $\frac{1}{r_{X_{m+2}}+b_{X_{m+2}}}$, then obtain the next state
\ELSIF{$S_t$ is in the form $(X_m, X_{m+1}, Y_{m+2})$}
\STATE Uniformly pick a red node which is adjacent to $Y_{m+2}$ in red level by chance $\frac{1}{r_{Y_{m+2}}+b_{X_{m+1}}}$, then obtain the next state
\STATE Or uniformly pick a blue node which is adjacent to $X_{m+1}$ by chance $\frac{1}{r_{Y_{m+2}}+b_{X_{m+1}}}$, then obtain the next state
\ELSE
\STATE ($S_t$ is in the form $(X_m, Y_{m+1}, Y_{m+2})$)
\STATE Uniformly pick 2 blue nodes $X_{m-1}$ and $X_{m+1}$ which are adjacent to $X_m$, and obtain the next state
\ENDIF 
\ENDWHILE
\STATE $\hat {|d_i|} = \frac{\hat {|C_i|}}{\sum_{j=1}^{14}\hat {|C_i|}}$, $i \in \{1,2,...,14\}$\\
\end{algorithmic}
\end{algorithm}.

\section{Experimental Evaluation}\label{sec:experiment}
We evaluate the performance of the proposed algorithms (\textsf{RWNbN}, \textsf{RWEbE}, \textsf{RWOMRN} and \textsf{RWMix}) on a set of multi-layer networks. 
The baseline algorithm is the random walk that can traverse both layers without restrictions (\textsf{RWNR}), and is thus statistically more 
accurate than the proposed algorithms. We claim that our proposed algorithms are effective if they yield the comparable 
accuracy as the baseline.  The algorithms are implemented in Python and we run experiments on a Linux machine with Intel 3.70 GHz CPU and 8G memory. Our purpose is to answer the following questions:

\begin{itemize}

\item Q1: How does the heterogeneity in the structure of different layers influence the accuracy of sampling?

\item Q2: How accurate are our sampling algorithms in synthetic and real-world two-layer graphs?

\item Q3: Can we balance the sampling accuracies of different types of graphlets?

\end{itemize}

\subsection{Experimental Setup}

We concatenate two single-layer networks into a two-layer multiplex network under two scenarios: 
i) both layers are \emph{synthetic} networks, ii) one layer is from the real-world and the other layer is synthetic. 
With the synthetic two-layer networks, we are able to qualitatively explore the relationship between the sampling accuracy and 
the network structure; with the real-world and syntetic OSNs, we can evaluate the accuracy of our algorithms in the wild.

\noindent\textbf{Error Metrics.} We consider two commonly used metrics to characterize the errors of graph sampling. 

\begin{itemize}
	\item \emph{Mean of relative error (MRE)} is defined as:  $E|\frac{|\hat{|C_i|}-|C_i||}{|C_i|}|$ over 1000 independent runs. It is 
	used to measure the closeness of estimation to the ground truth. 
	
	\item \emph{Normalized root mean square error (NRMSE)} is defined as: $NRMSE(\hat{|C_i|})=\frac{\sqrt{E[(\hat{|C_i|}-|C_i|)^2]}}{|C_i|}$ =$\frac{\sqrt{Var[\hat{|C_i|}]+E[(\hat{|C_i|}-|C_i|)^2]}}{|C_i|}$. It is used to measure the variance and bias of the 
	estimators jointly.

\end{itemize}

The number of random walk steps is set to 20k that only visits a small fraction of the nodes. The number of independent runs is 
set to 1000 in order to show the convergence of $NRMSE$ as the number of random walks progresses. 

\subsection{Experiment Results}

Qualitative and quantitative analyses are presented in this section. The former is based on experiments on synthetic datasets while the latter corresponds to the real and synthetic datasets.

\textbf{Qualitative analysis on synthetic graphs.} 

In an attempt to study the impact of various factors on the sampling performance, we run experiments on several synthetic multiplex networks with two generated layers. To be specific, we focus on two factors: degree distribution of the network per layer and the connection of two layers.

The blue layer is generated from three categories: ER (Erdos-Renyi random graph), SW (small world graph) and BA (Barabasi-Albert scale-free graph) while the red layer is generated to balance the concentrations of all graphlets. For each category, the two layers are interconnected by three ways. The first is that every node in the blue layer has a corresponding node in red layer (i.e. all the nodes have a one-to-one mapping). The second is that the blue layer and the red layer are of the same scale (i.e. the same number of nodes), while a half of nodes in the blue layer connect to their counterparts in the red layer. The third is 
that the blue graph is two times larger than the 
red graph, and all the red nodes connect to the 
half of the nodes in the blue graph. 
Note that we cannot exhaust all the possible combinations of 
two layers, and we believe that the sampling is meaningful 
if the sizes of two layers are not too much different. 

We denote these three two-layer graphs by \#1, \#2, \#3.
Then, ER1 (resp. ER2, ER3) indicates \#1 (resp. \#2, \#3) Erdos-Renyi graph, and the other notations are the same.
We summarize the basic information of those nine datasets in 
Table \ref{tab7}, and enumerate the ground-truth concentrations of some representative graphlets. For example, the number \#8 
refers to the $8^{th}$ graphlet. One can observe that the
ground-truth concentrations of the graphlets are highly 
diverse, and a rare graphlet is believed to have a high 
estimation error, and vice versa. In this set of 
experiments, we mainly focus on the relative change that 
our algorithms perform in different datasets, rather than 
the absolute magnitudes of the estimation errors.

\begin{table}[!htb]
\centering
\caption{Information of data sets }
\label{tab7}
\begin{tabular}{|c|c|c|c|c|c|c|}
\hline
Graph&$|\mathcal{V}|$&$|\mathcal{E}|$&\#8($10^{-6}$)&\#10($10^{-6}$)&\#12($10^{-6}$)\\ 
\hline
ER1&100K&948K&8.06&5.27&5.67\\
\hline
ER2&100K&753K&3.97&6.7&2.11\\
\hline
ER3&100K&753K&3.85&4.97&1.86\\
\hline
SW1&100K&757K&7.88&3.78&1764\\
\hline
SW2&120K&824K&961&413&1843\\
\hline
SW3&100K&651K&7.37&10.2&1800\\
\hline
BA1&100K&946K&3.25&2.06&54\\
\hline
BA2&120K&1068K&12.7&2.16&44\\
\hline
BA3&100K&957K&38.5&24.9&32.8\\
\hline
\end{tabular}
\end{table}

\begin{figure}[!h!t]
\begin{minipage}[!t]{0.5\textwidth}
\centering
\includegraphics[width=3in]{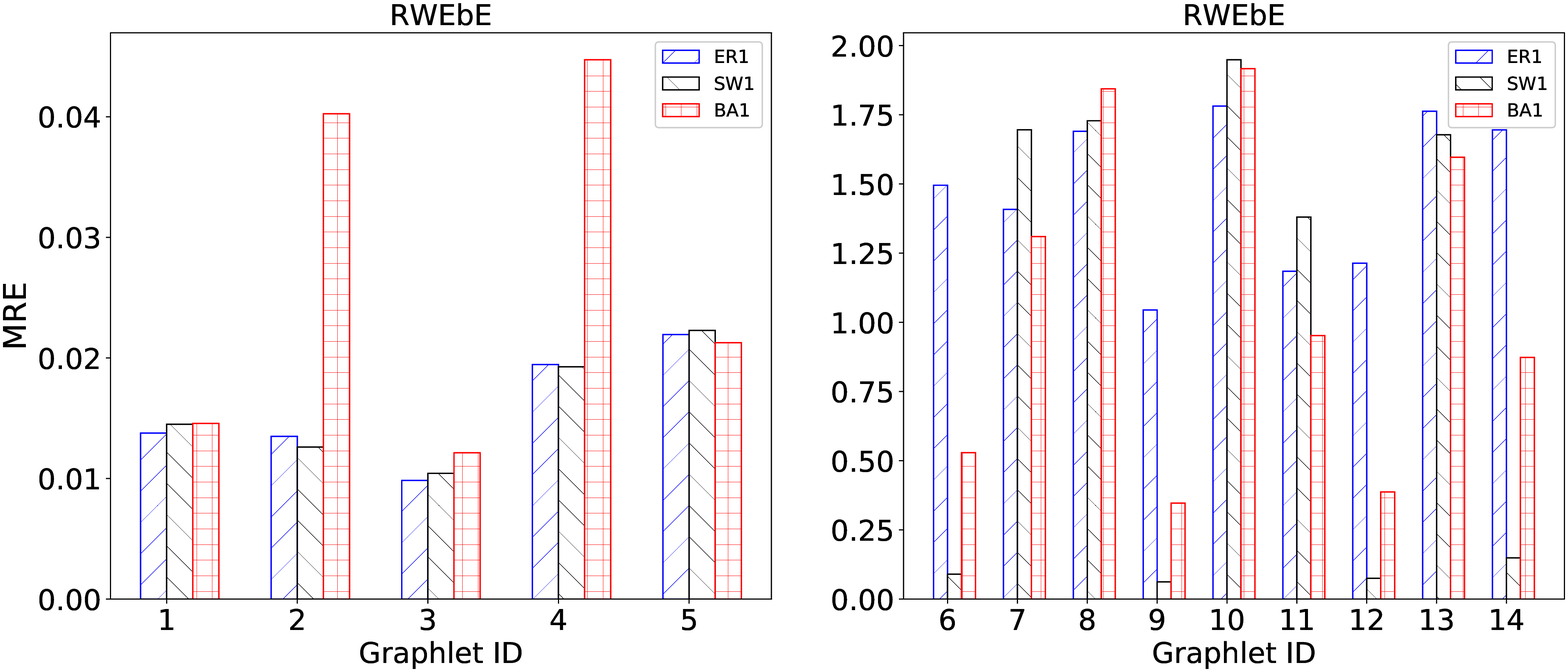}
\caption{\textbf{MRE on ER1, SW1 and BA1}}
\label{Results on 3 1 graphs}
\end{minipage}%
\begin{minipage}[!t]{0.5\textwidth}
\centering
\includegraphics[width=3in]{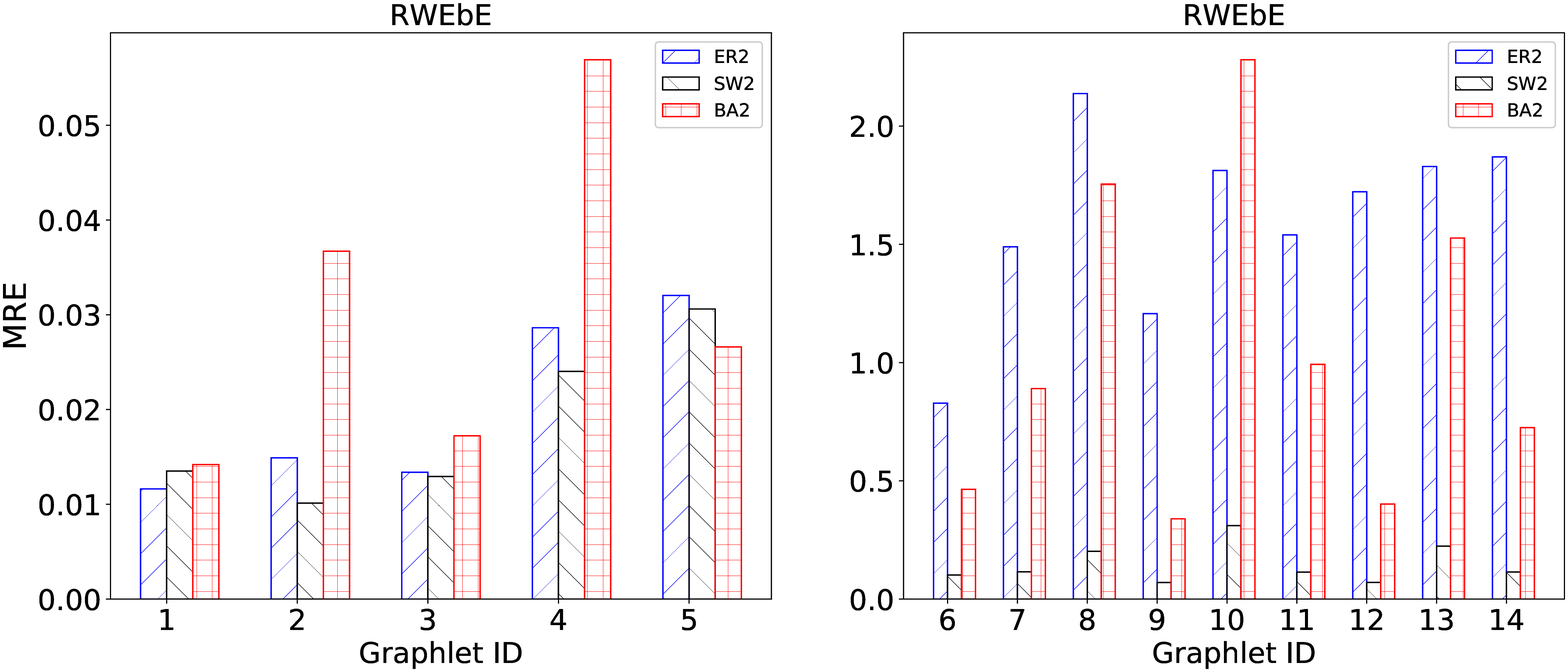}
\caption{\textbf{MRE on ER2, SW2 and BA2}}
\label{Results on 3 2 graphs}
\end{minipage}
\end{figure}

\begin{figure}[!h!t]
\begin{minipage}[t]{0.5\textwidth}
\centering
\includegraphics[width=3in]{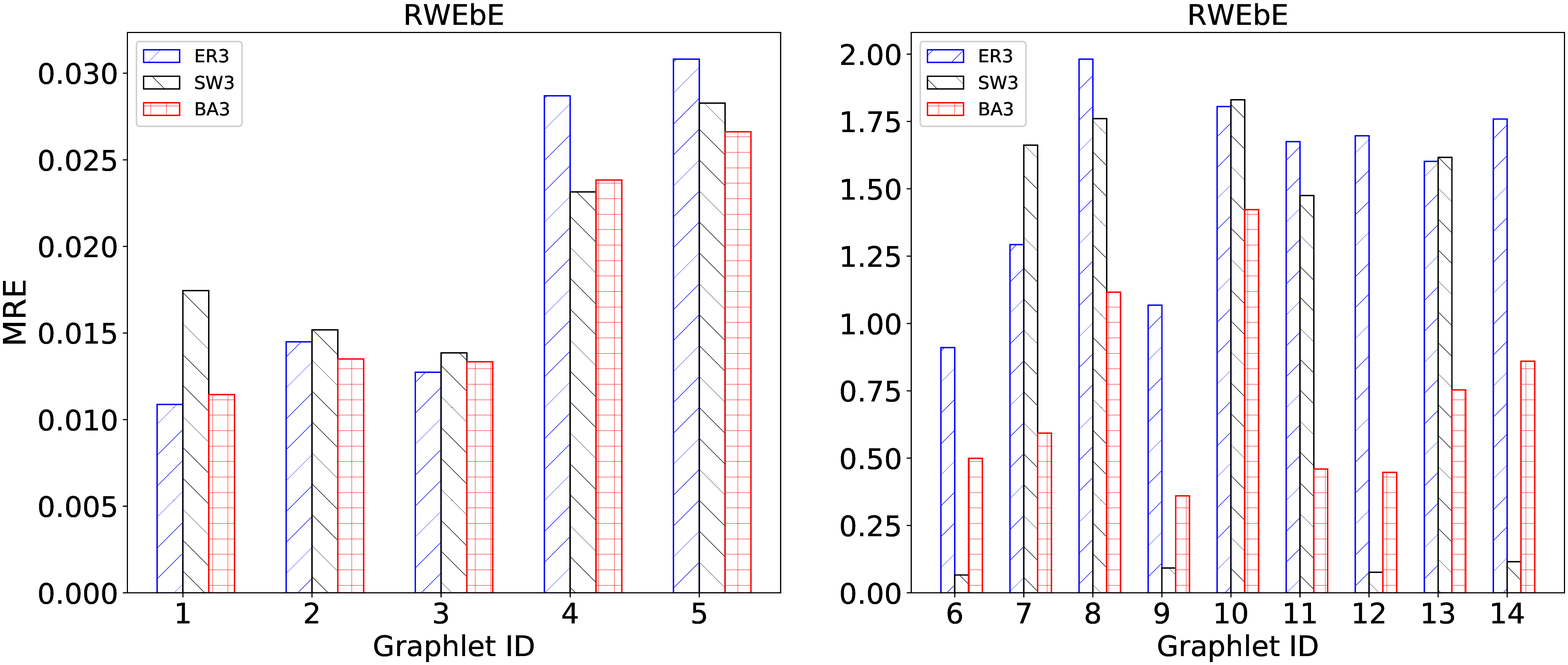}
\caption{\textbf{MRE on ER3, SW3 and BA3}}
\label{Results on 3 3 graphs}
\end{minipage}%
\begin{minipage}[t]{0.5\textwidth}
\centering
\includegraphics[width=3in]{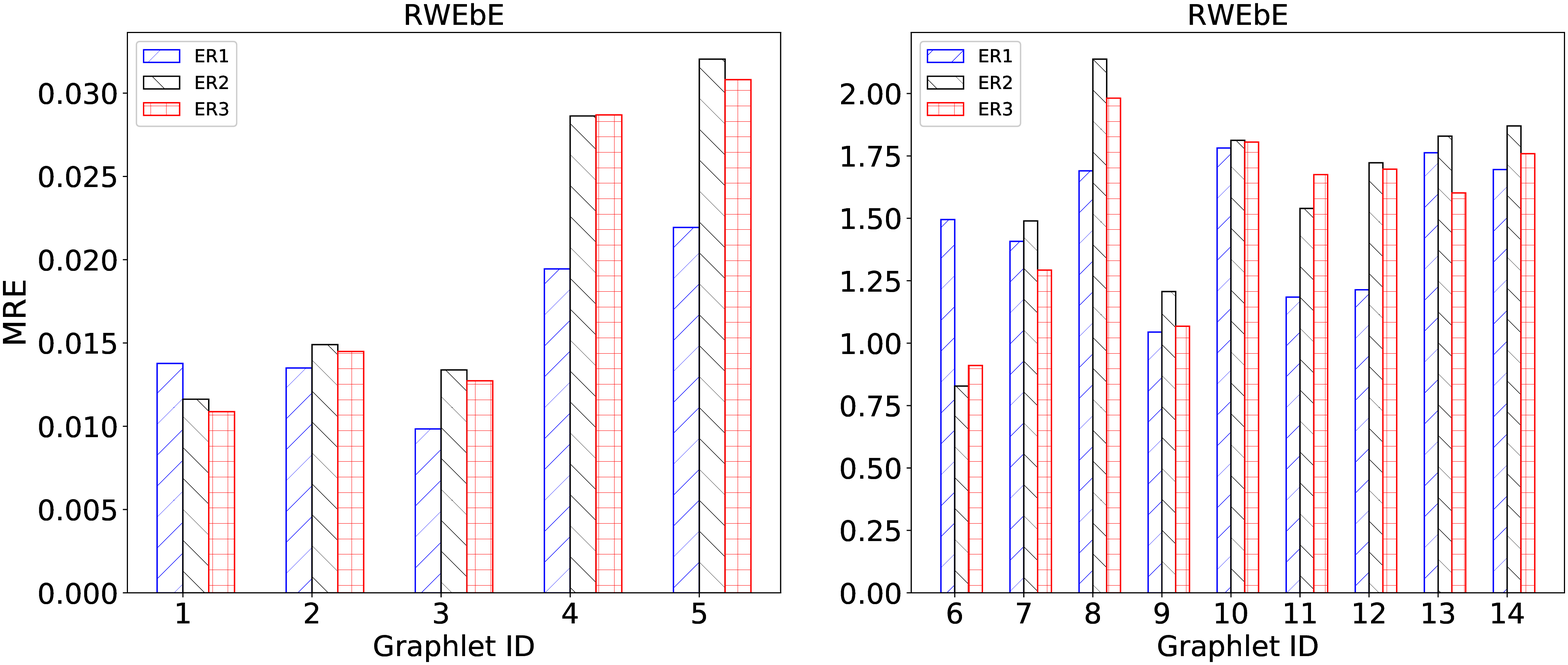}
\caption{\textbf{MRE on ER1, ER2 and ER3}}
\label{Results on 3 ER graphs}
\end{minipage}
\end{figure}

\begin{figure}[!h!t]
\begin{minipage}[t]{0.5\textwidth}
\centering
\includegraphics[width=3in]{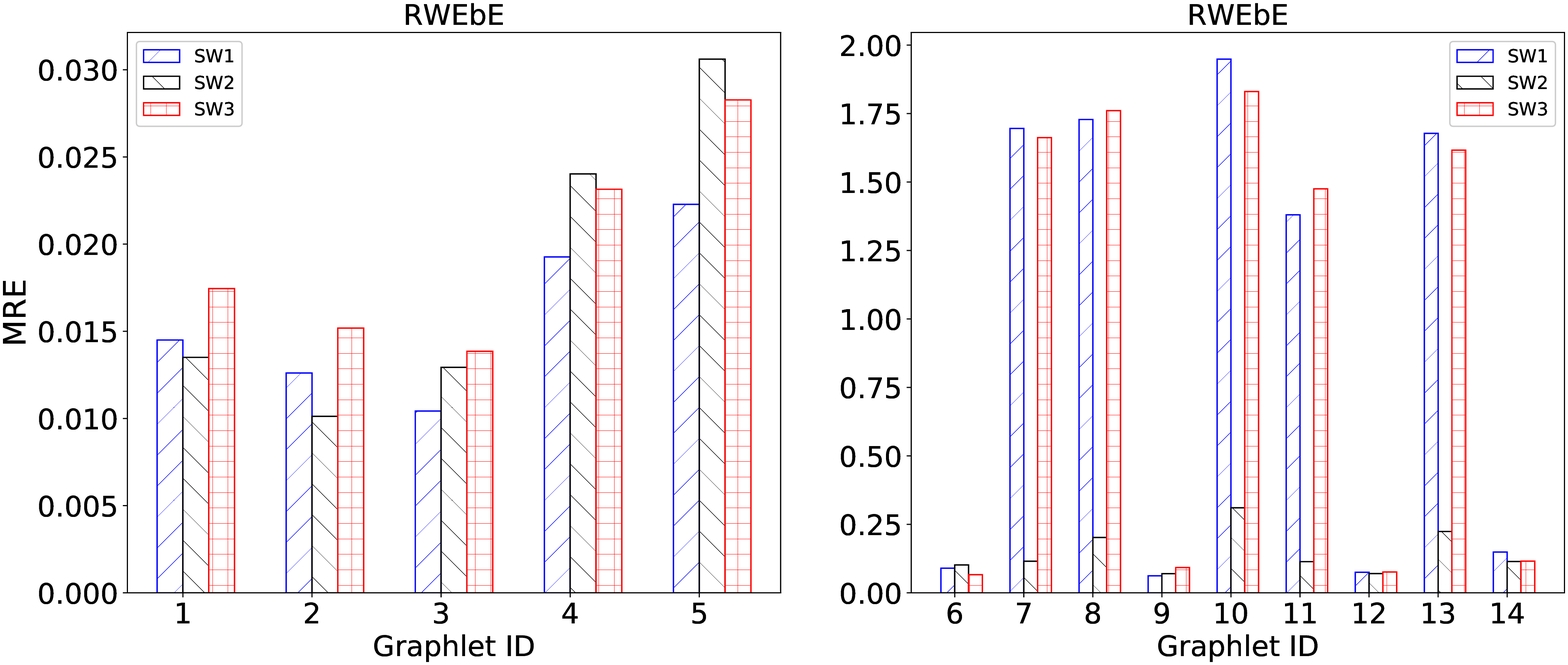}
\caption{\textbf{MRE on SW1, SW2 and SW3}}
\label{Results on 3 SW graphs}
\end{minipage}%
\begin{minipage}[t]{0.5\textwidth}
\centering
\includegraphics[width=3in]{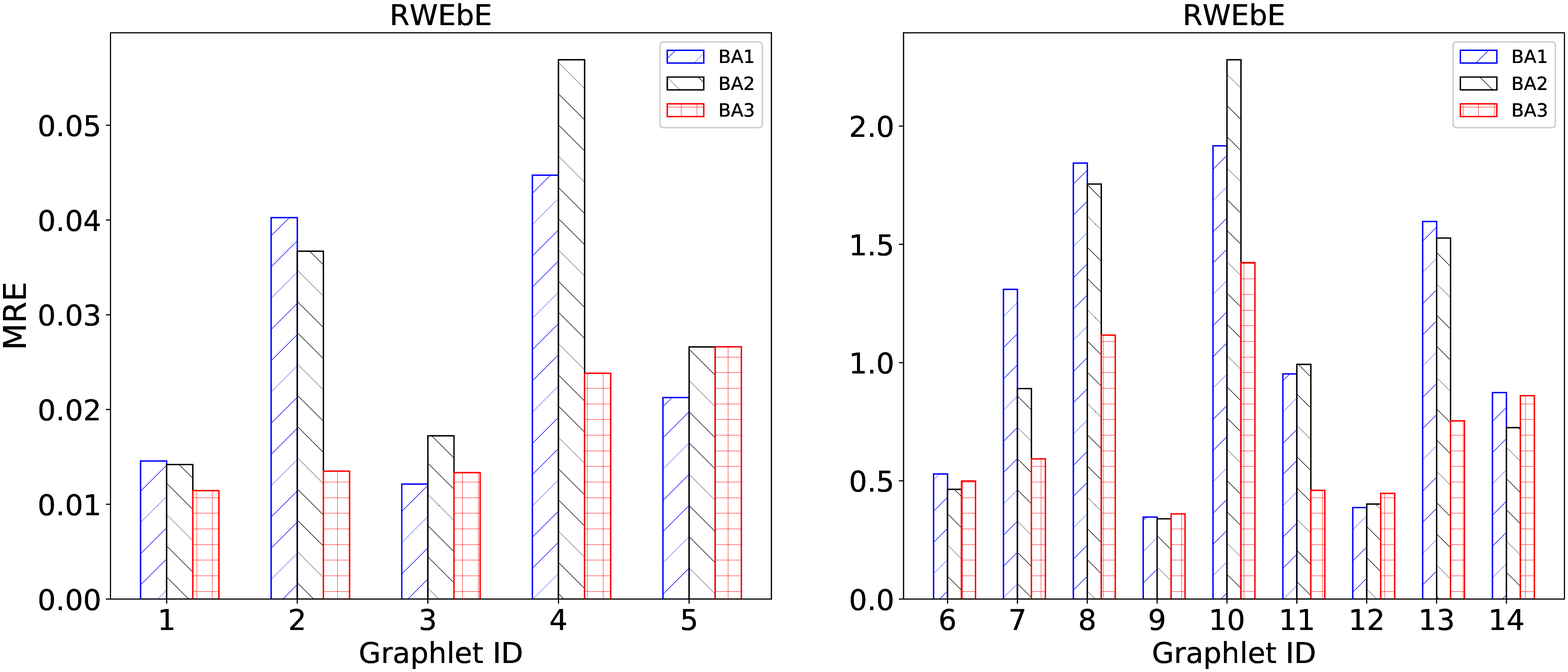}
\caption{\textbf{MRE on BA1, BA2  and BA3}}
\label{Results on 3 BA graphs}
\end{minipage}
\end{figure}

In what follows, we fix the sampling algorithm while changing the graph topology. 
Due to lack of space, we stick to \textsf{RWEbE}, and similar conclusions can be drawn from other sampling algorithms. 
Firstly, the network topology does influence the accuracy of our sampling algorithm, but through the concentration of graphlets. 
Fig. \ref{Results on 3 1 graphs}, Fig. \ref{Results on 3 2 graphs} and Fig. \ref{Results on 3 3 graphs} show the MREs 
of different graphlets on ER, SW and BA networks, respectively. 
The MSEs of the graphlets $\{1, 2, 3, 4, 5\}$ are always at the order of $10^{-2}$, indicating the high sampling accuracy. 
Those of the graphlets $\{6,7,8,9,10,11,12,13,14\}$ are relatively higher which are less accurately estimated. 
The decrease of sampling accuracy is originated from the very low graphlet concentration (as shown in the Appendix). 
In Fig. \ref{Results on 3 1 graphs}, one can see that the MREs of graphlets $\{2,4\}$ 
are obviously higher on BA1 while the MREs of graphlets \{6,9,12,14\} are much lower on SW1. 
Partial reasons attribute to that each node in SW1 has more uniform number of edges than ER1 and BA1, thus 
the rarest graphlets still have higher concentrations compared with ER1 and BA1. 
However, there is no graph that has the advantage on every graphlet. Similar observations can be strengthened 
in Fig. \ref{Results on 3 2 graphs} and Fig. \ref{Results on 3 3 graphs}.

\begin{figure}[!h!t]
\begin{minipage}[!h!t]{0.33\textwidth}
\centering
\includegraphics[width=2in]{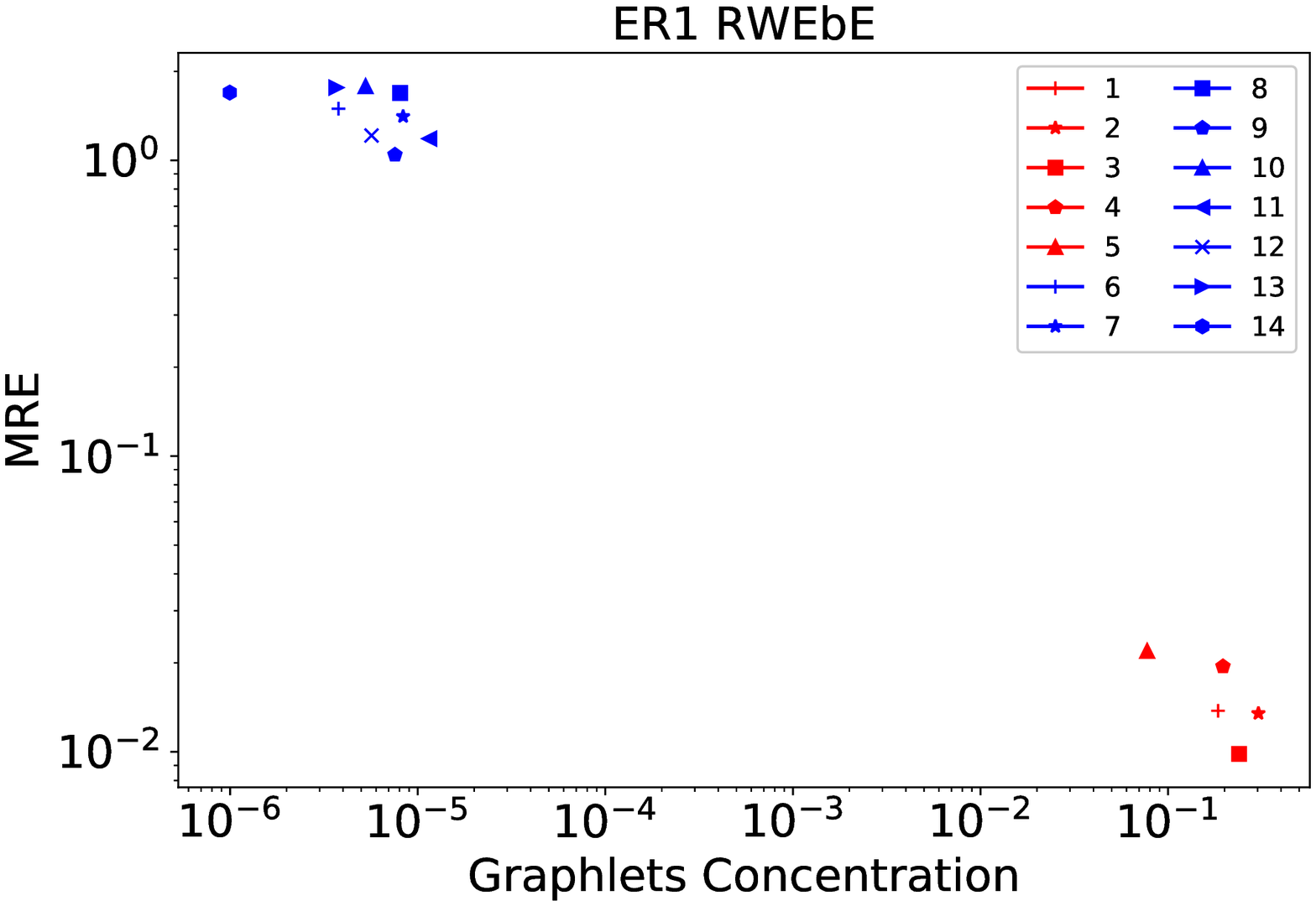}
\caption{Graphlet Ratio vs MRE on ER1}
\label{Relationship between concentration and MRE on ER1}
\end{minipage}%
\begin{minipage}[!h!t]{0.33\textwidth}
\centering
\includegraphics[width=2in]{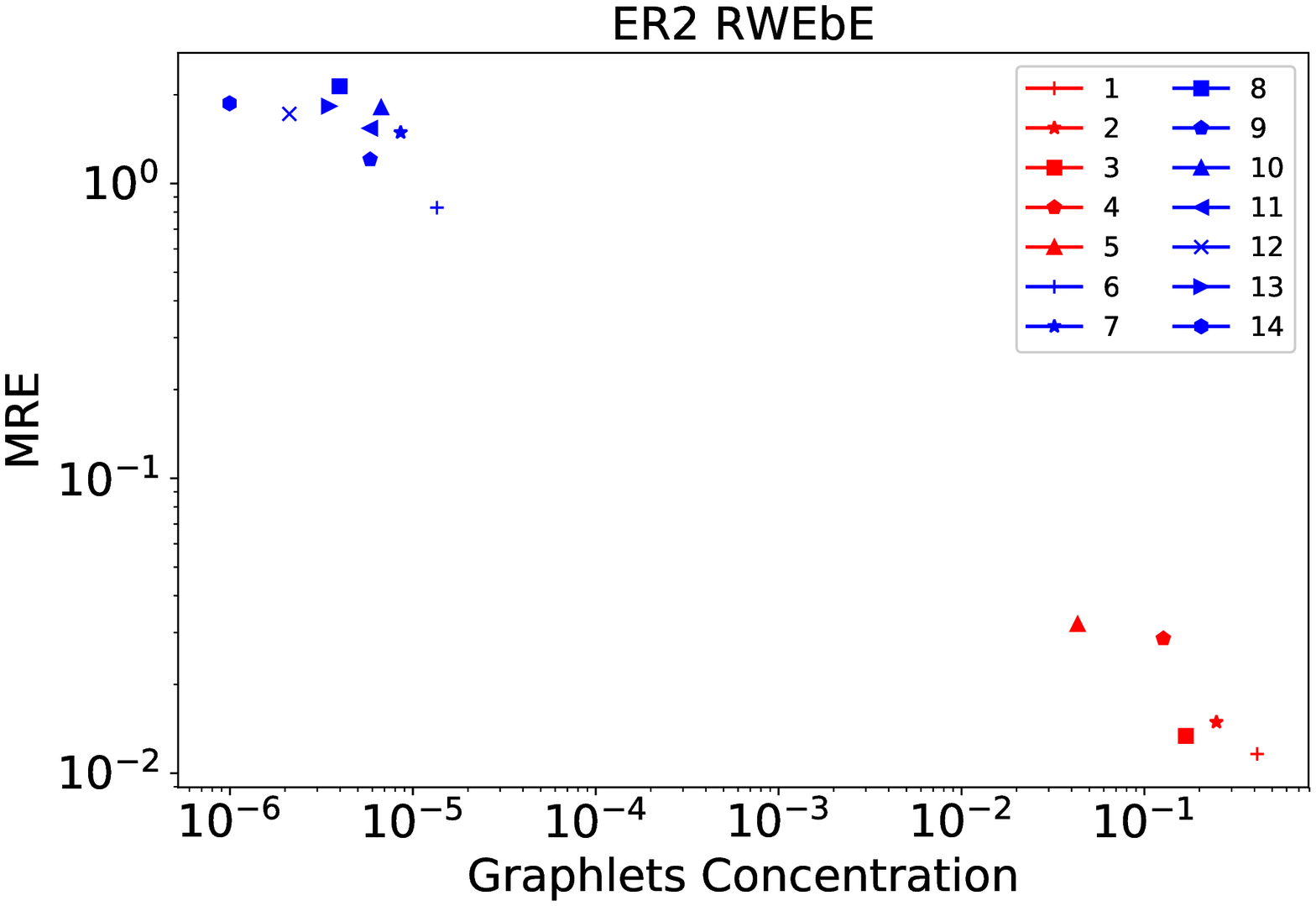}
\caption{Graphlet Ratio vs MRE on ER2}
\label{Relationship between concentration and MRE on ER2}
\end{minipage}
\begin{minipage}[!h!t]{0.33\textwidth}
\centering
\includegraphics[width=2in]{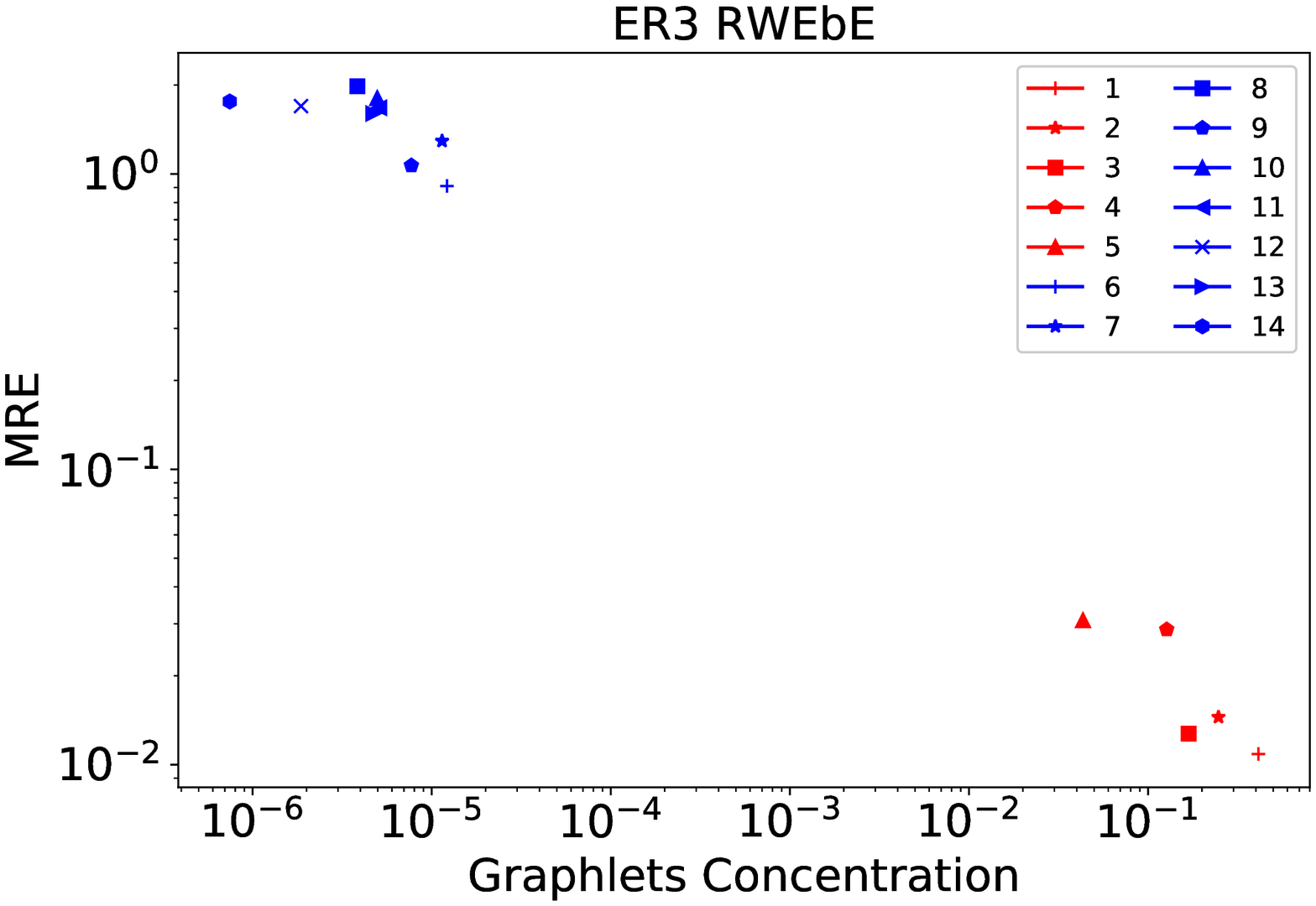}
\caption{Graphlet Ratio vs MRE on ER3}
\label{Relationship between concentration and MRE on ER3}
\end{minipage}%
\end{figure}

\begin{figure}[!h!t]
\begin{minipage}[!h!t]{0.33\textwidth}
\centering
\includegraphics[width=2in]{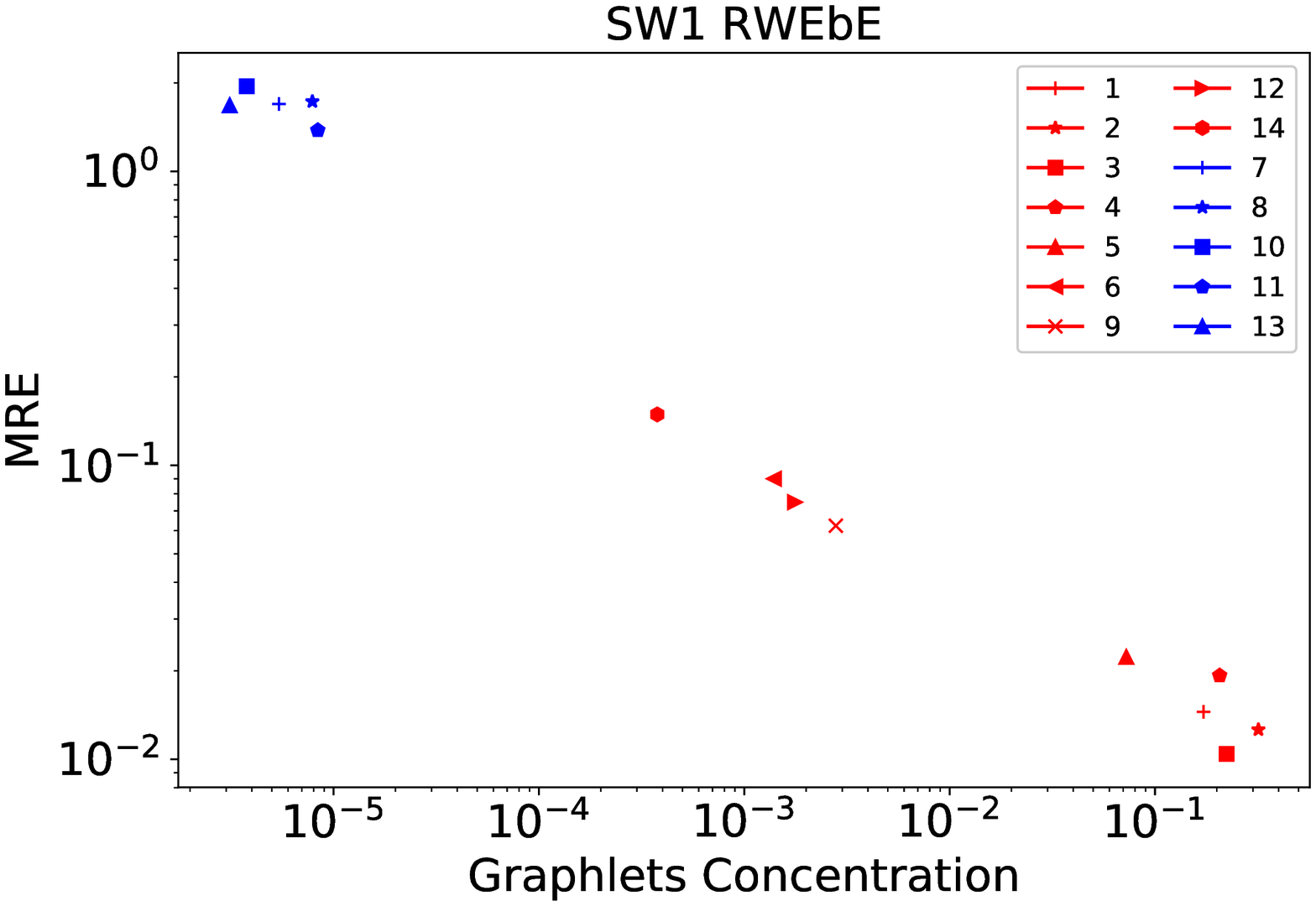}
\caption{Graphlet Ratio vs MRE on SW1}
\label{Relationship between concentration and MRE on SW1}
\end{minipage}
\begin{minipage}[!h!t]{0.33\textwidth}
\centering
\includegraphics[width=2in]{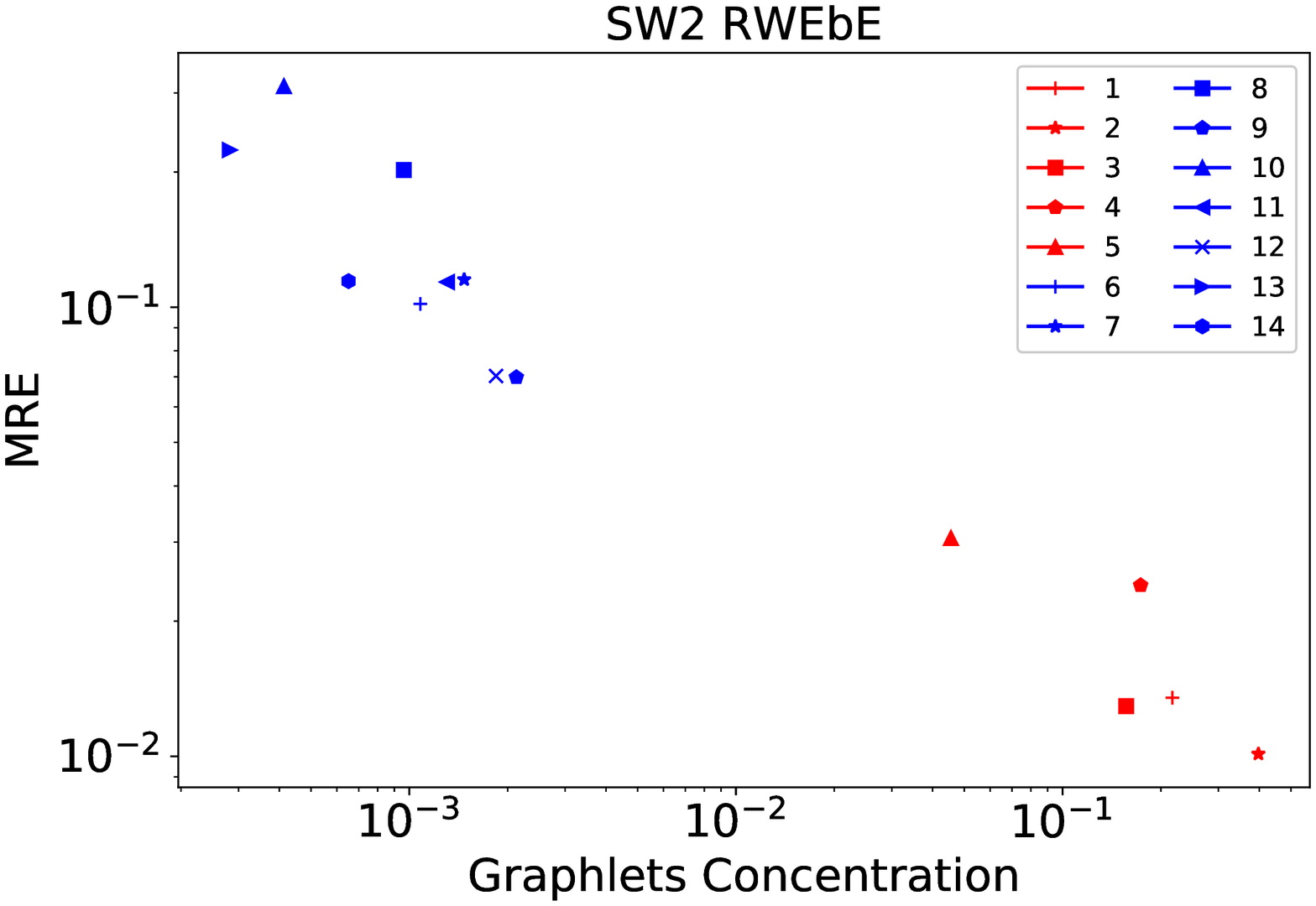}
\caption{Graphlet Ratio vs MRE on SW2}
\label{Relationship between concentration and MRE on SW2}
\end{minipage}%
\begin{minipage}[!h!t]{0.33\textwidth}
\centering
\includegraphics[width=2in]{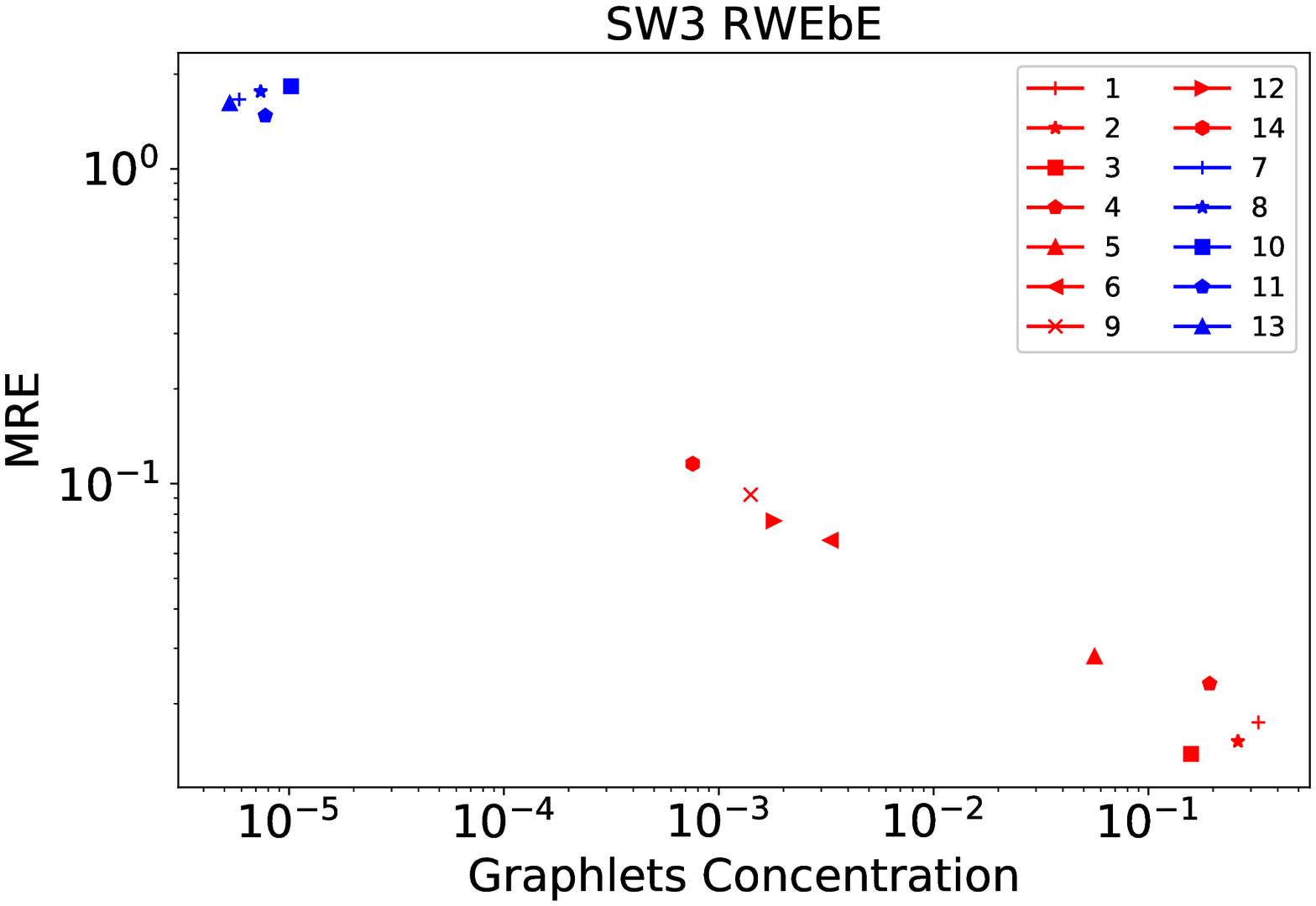}
\caption{Graphlet Ratio vs MRE on SW3}
\label{Relationship between concentration and MRE on SW3}
\end{minipage}
\end{figure}

\begin{figure}[!h!t]
\begin{minipage}[!h!t]{0.33\textwidth}
\centering
\includegraphics[width=2in]{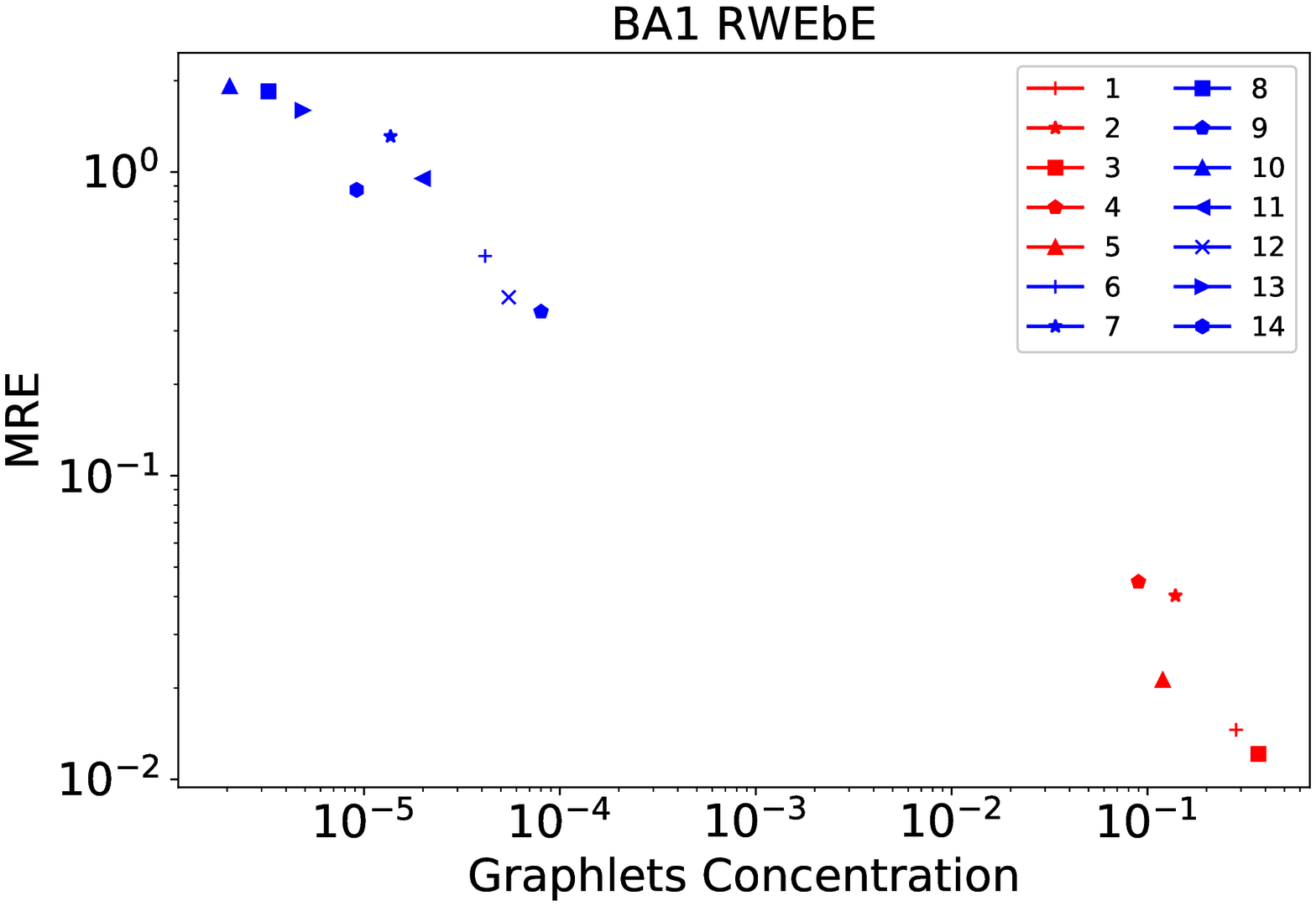}
\caption{Graphlet Ratio vs MRE on BA1}
\label{Relationship between concentration and MRE on BA1}
\end{minipage}%
\begin{minipage}[!h!t]{0.33\textwidth}
\centering
\includegraphics[width=2in]{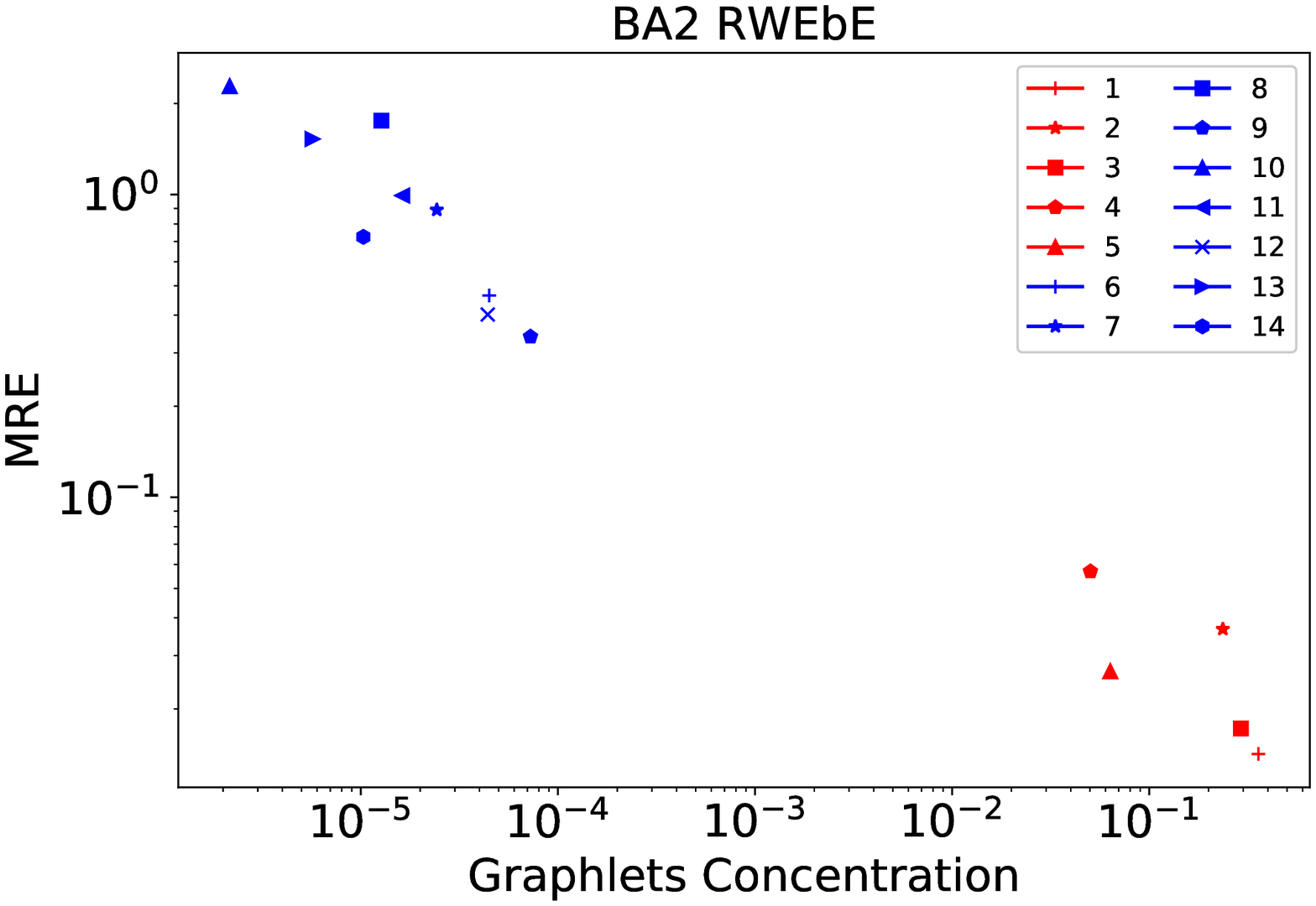}
\caption{Graphlet Ratio vs MRE on BA2}
\label{Relationship between concentration and MRE on BA2}
\end{minipage}
\begin{minipage}[!h!t]{0.33\textwidth}
\centering
\includegraphics[width=2in]{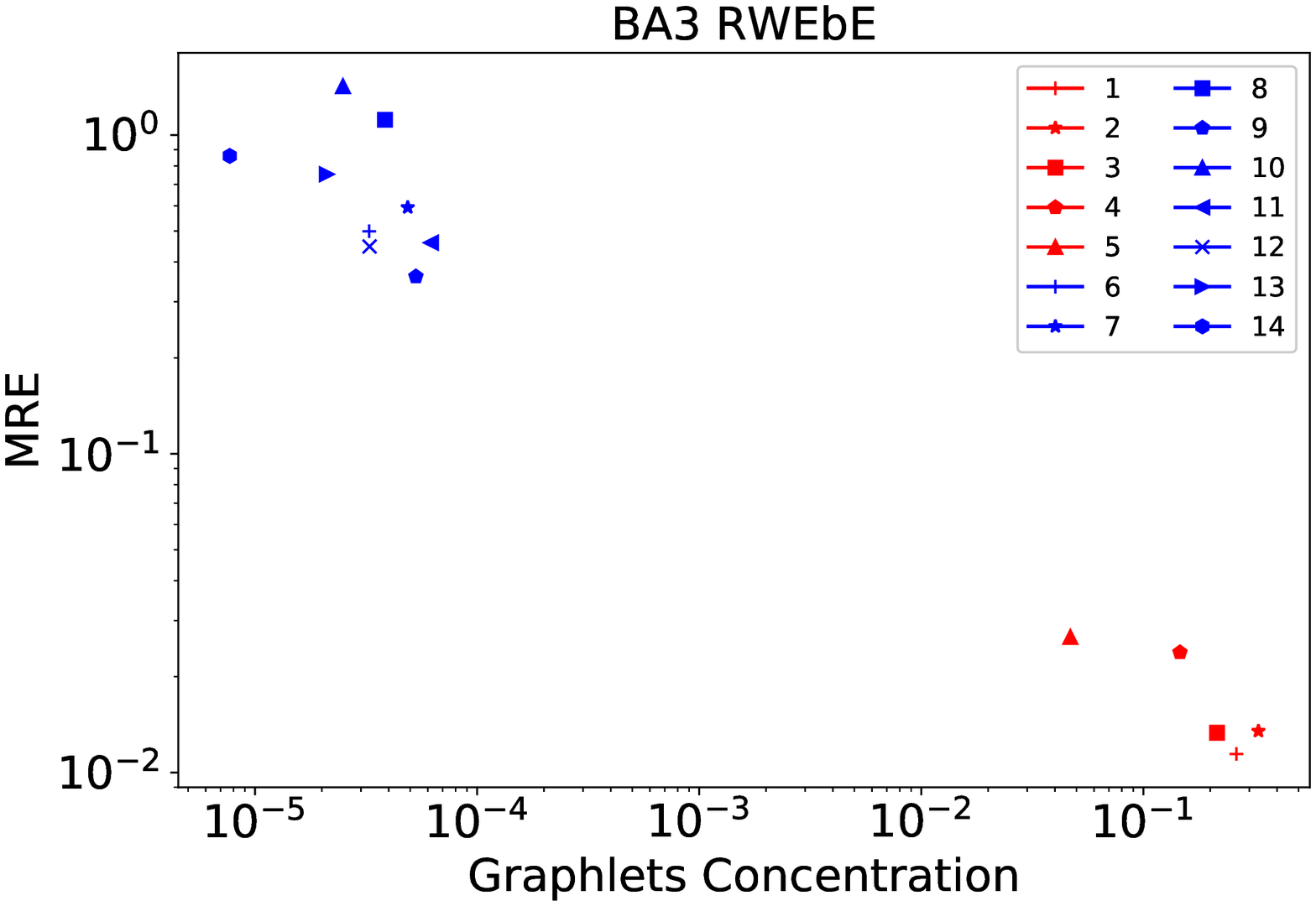}
\caption{Graphlet Ratio vs MRE on BA3}
\label{Relationship between concentration and MRE on BA3}
\end{minipage}%
\end{figure}

Secondly, changing the density of inter-layer links usually does not influence the performance of our sampling algorithm. 
Fig. \ref{Results on 3 ER graphs} evaluates the MREs of three ER graphs in which their differences are marginal. 
In Fig. \ref{Results on 3 SW graphs}, the MREs of graphlets $\{7,8,10,11,13\}$ on SW2 are much lower than on 
SW1 and SW3. In Fig. \ref{Results on 3 BA graphs}, the MREs of graphlets $\{7,8,10,11,13\}$. 
We hereby argue that the extremely low concentration of graphlets, instead of the sampling algorithm itself, causes the 
increased mean of relative error. The relationships between the ground-truth graphlet concentration and the MRE on all
 the graphs are shown in Fig. \ref{Relationship between concentration and MRE on ER1}$\sim$
 Fig. \ref{Relationship between concentration and MRE on BA3} where the $x$-coordinate indicates the logarithm of the concentration ratio 
 and the $y$-coordinate is the logarithm of the MRE. 
One can observe that all the rare graphlets (from $6^{th}$ to $14^{th}$ at the order of $10^{-6}$) encounter relatively large MREs on every network. 
 As the concentration ratio increases, the MRE decreases accordingly. 
In Fig. \ref{Relationship between concentration and MRE on SW2}, the estimation on SW2 is more accurate simply because that the rare graphlets have much higher concentration ratios 
(at the order of $10^{-4}$) than the other networks. 
Therefore, we can conclude that the accuracy of our sampling algorithm is throttled by the ground-truth concentration ratio of 
a graphlet.

\textbf{Quantitative analysis with real-world graphs.} 

We evaluate the accuracy and convergence of the proposed algorithms with real world graph datasets. 
The blue layer comes from the real-world, and the red layer is 
synthetic because there are few large-scale two-layer network datasets publicly available. Table \ref{tab8} lists the basic information of these datasets including Epinions and Facebook.

\begin{table}[!htb]
\centering
\caption{Information of data sets }
\label{tab8}
\begin{tabular}{|c|c|c|c|c|c|}
\hline
Graph&$|\mathcal{V}|$&$|\mathcal{E}|$&\#7($10^{-6}$)&\#9($10^{-6}$)&\#11($10^{-6}$)\\ 
\hline
Epinions1&76K&842K&23&8577&40\\
\hline
Epinions2&76K&990K&20&6410&53\\
\hline
Facebook1&63K&1532K&55&18038&68\\
\hline
Facebook2&63K&1841K&43&13269&96\\
\hline
\end{tabular}
\end{table}

\begin{figure}[!h!t]
\begin{minipage}[!h!t]{0.5\textwidth}
\centering
\includegraphics[width=3.3in,height=1.5in]{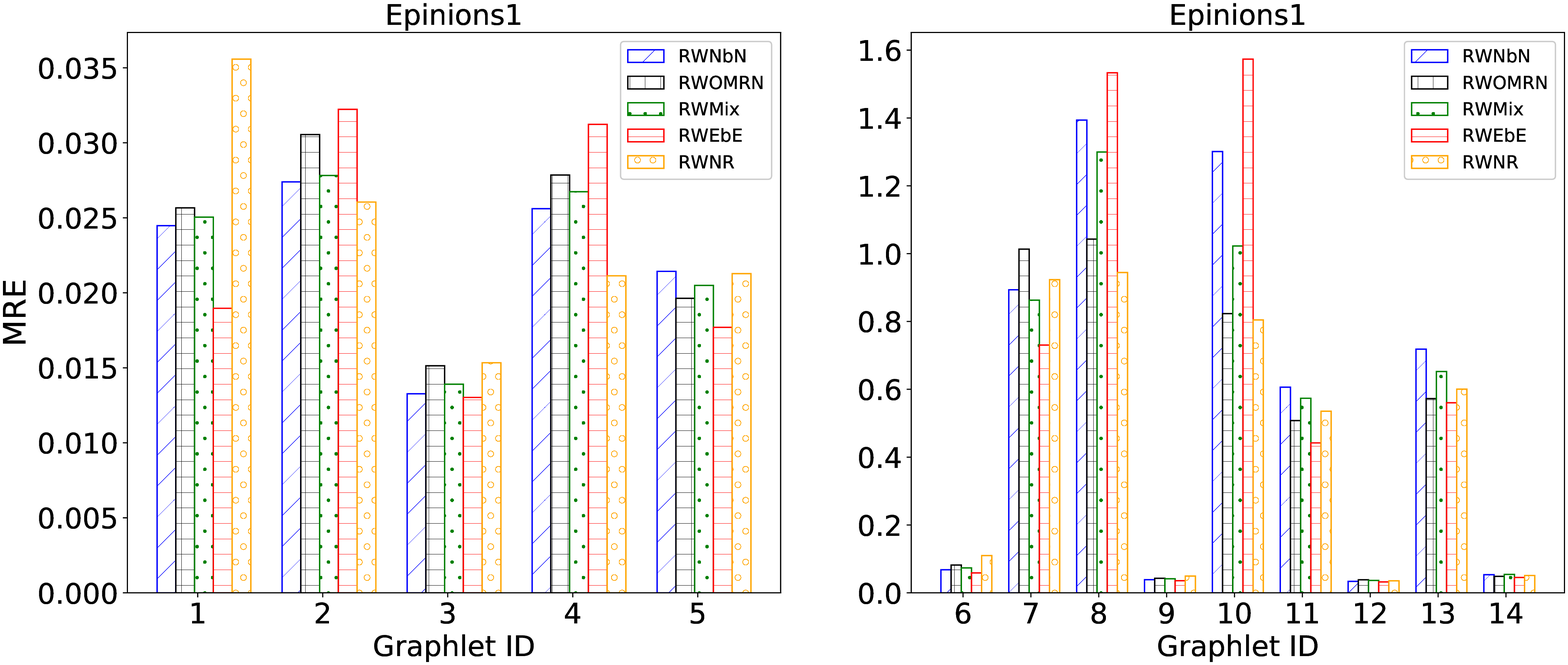}
\caption{MRE on Epinions1}
\label{MRE on Epinions1}
\end{minipage}%
\begin{minipage}[!h!t]{0.5\textwidth}
\centering
\includegraphics[width=3.3in,height=1.5in]{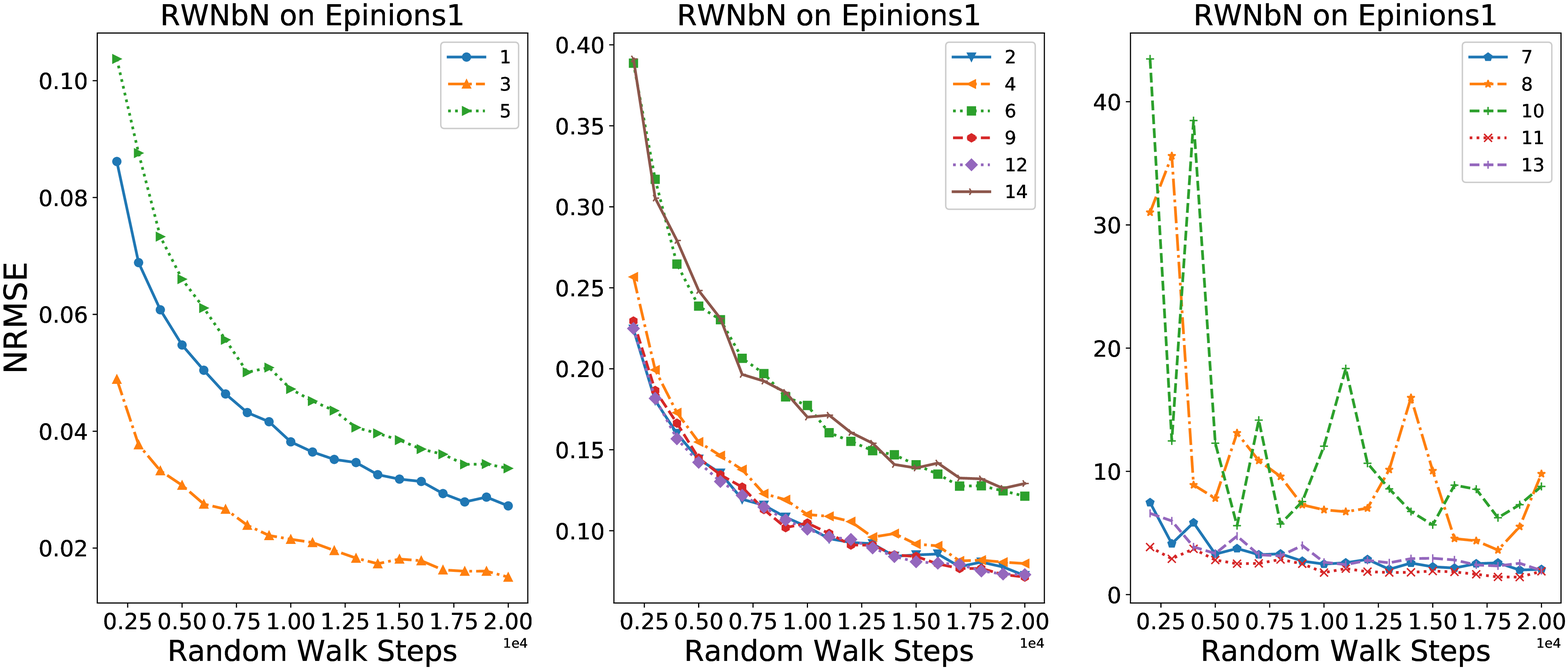}
\caption{RWNbN on Epinions1}
\label{RWNbN on Epinions1}
\end{minipage}
\end{figure}

Fig. \ref{MRE on Epinions1} shows the MRE of all the graphlets using the sampling approaches \textsf{RWNbN}, \textsf{RWOMRN}, \textsf{RWMiX}, \textsf{RWEbE} and \textsf{RWNR}. 
The MREs of the Epinions1 network are shown in Fig. \ref{MRE on Epinions1}. 
The first observation is that all the proposed algorithms achieve satisfactory accuracies on the set of graphlets $\{1, 2, 3, 4, 5, 6, 9, 12, 14\}$, with the MRE ranging from below 0.01 to around 0.1. To be specific, for graphlets with concentration around $10^{-3}$ ($\{6,9,12,14\}$), the MREs are about 0.1. For graphlets with concentration around $10^{-2}$ ($\{2,4\}$), the MREs are about 0.07. For graphlets with concentration higher than $10^{-2}$ ($\{1,3,5\}$), the MREs are around 0.02 - 0.03. The MREs on the set of graphlets $\{7, 8, 10, 11, 13\}$ 
are much higher due to their concentrations below $10^{-5}$. 
The second observation is that \textsf{RWNbN} has a better MRE than \textsf{RWOMRN} with more blue edges, and 
underperforms \textsf{RWOMRN} with more red edges, and the MRE of \textsf{RWMix} is usually in between. 
Hence, given a fixed sampling budget, we can balance the sampling accuracy of different types of graphlets by 
splitting this budget appropriately. As the third observation, the MRE of  \textsf{RWNbN}, \textsf{RWOMRN}, \textsf{RWMiX} and \textsf{RWEbE} is comparable to that of \textsf{RWNR} which allows the random walk on both layers. 
This implies that even though the random walk on the red graph is restricted, the sampling of two-layer graphlets can still be 
achieved with high accuracy. Last but not the least, no algorithm dominates the others on every graphlet.

\begin{figure}[!h!t]
\begin{minipage}[!h!t]{0.5\textwidth}
\centering
\includegraphics[width=3.3in,height=1.5in]{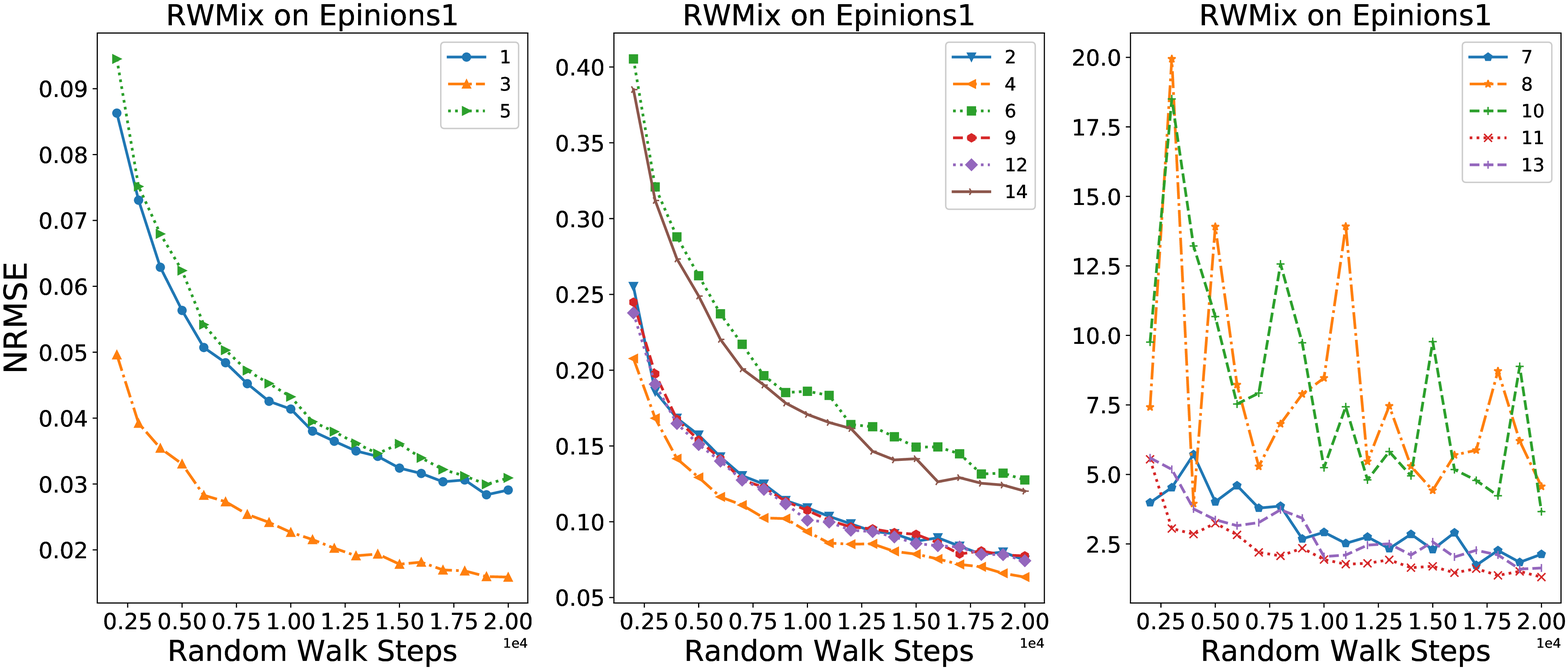}
\caption{RWMix on Epinions1}
\label{RWMix on Epinions1}
\end{minipage}%
\begin{minipage}[!h!t]{0.5\textwidth}
\centering
\includegraphics[width=3.3in,height=1.5in]{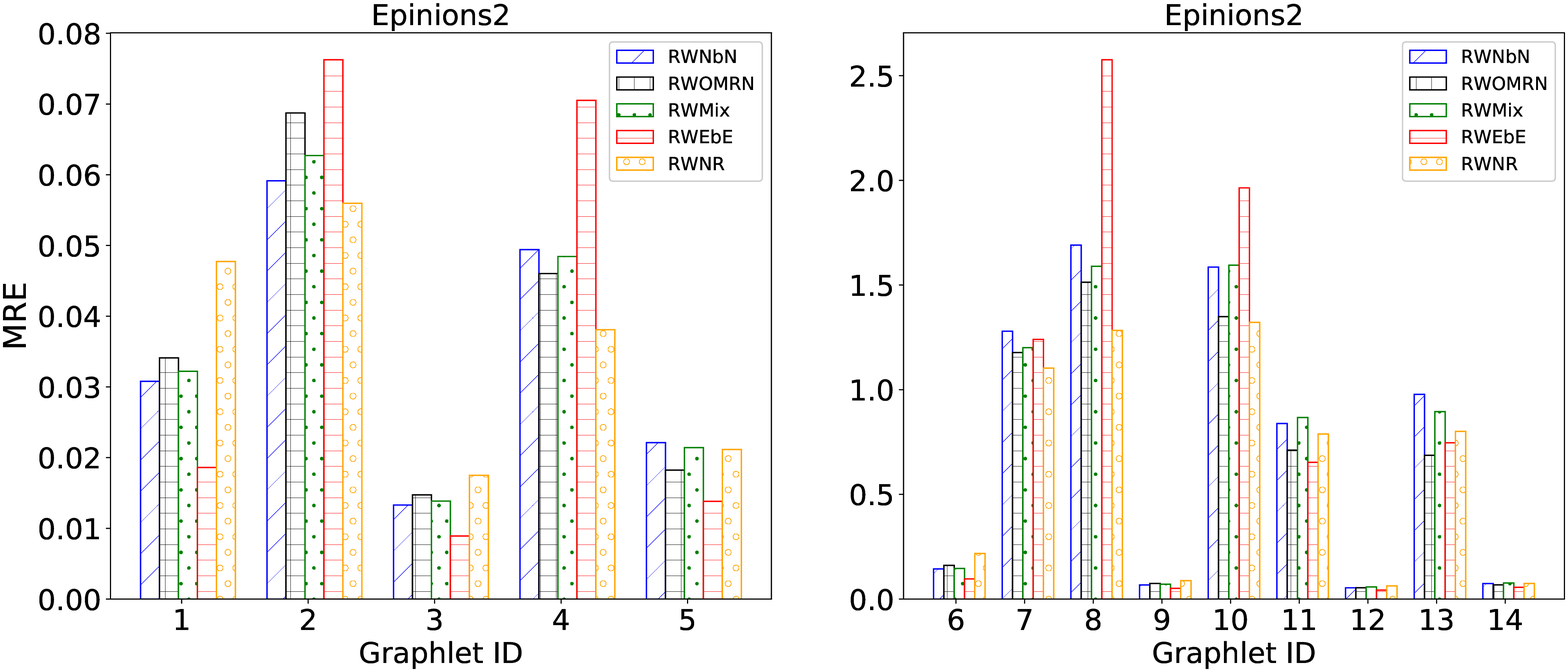}
\caption{MRE on Epinions2}
\label{MRE on Epinions2}
\end{minipage}
\end{figure}

\begin{figure}[!h!t]
\begin{minipage}[t]{0.5\textwidth}
\centering
\includegraphics[width=3.3in,height=1.5in]{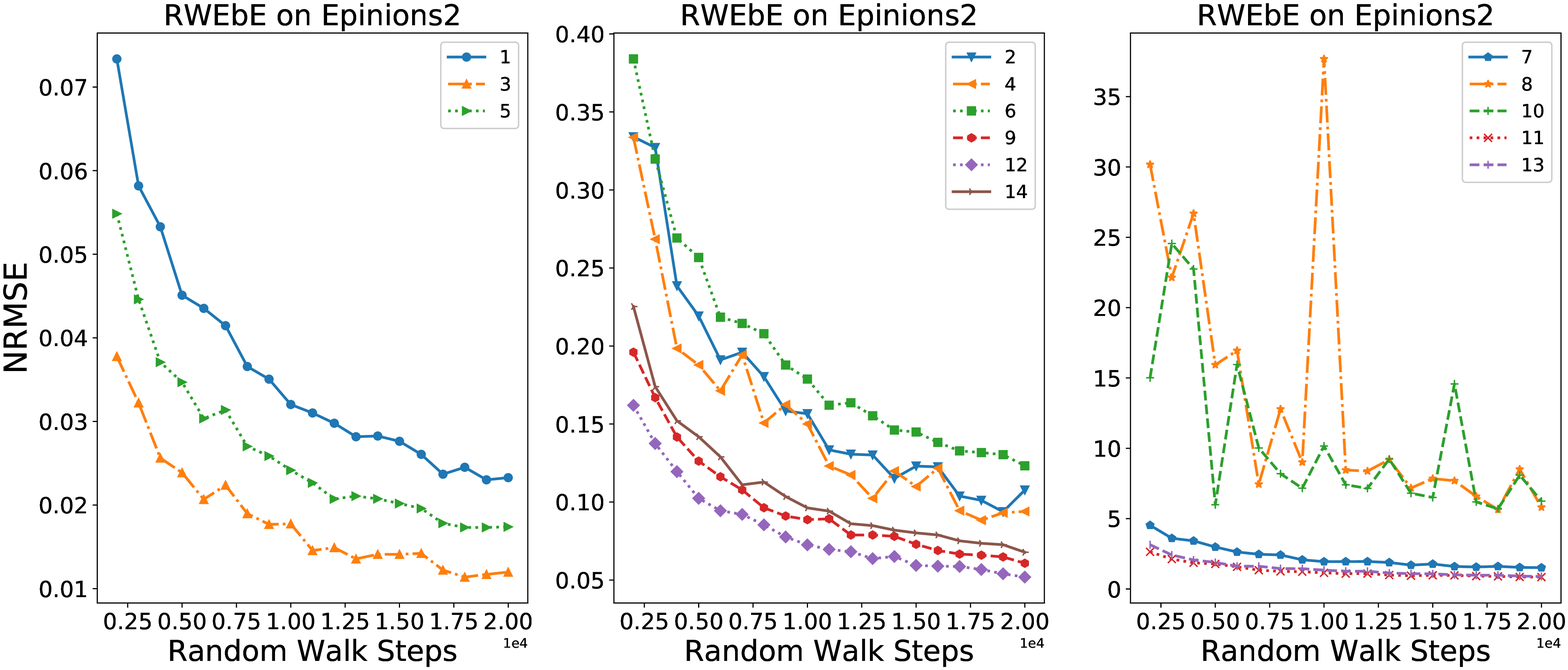}
\caption{RWEbE on Epinions2}
\label{RWEbE on Epinions2}
\end{minipage}%
\begin{minipage}[t]{0.5\textwidth}
\centering
\includegraphics[width=3.3in,height=1.5in]{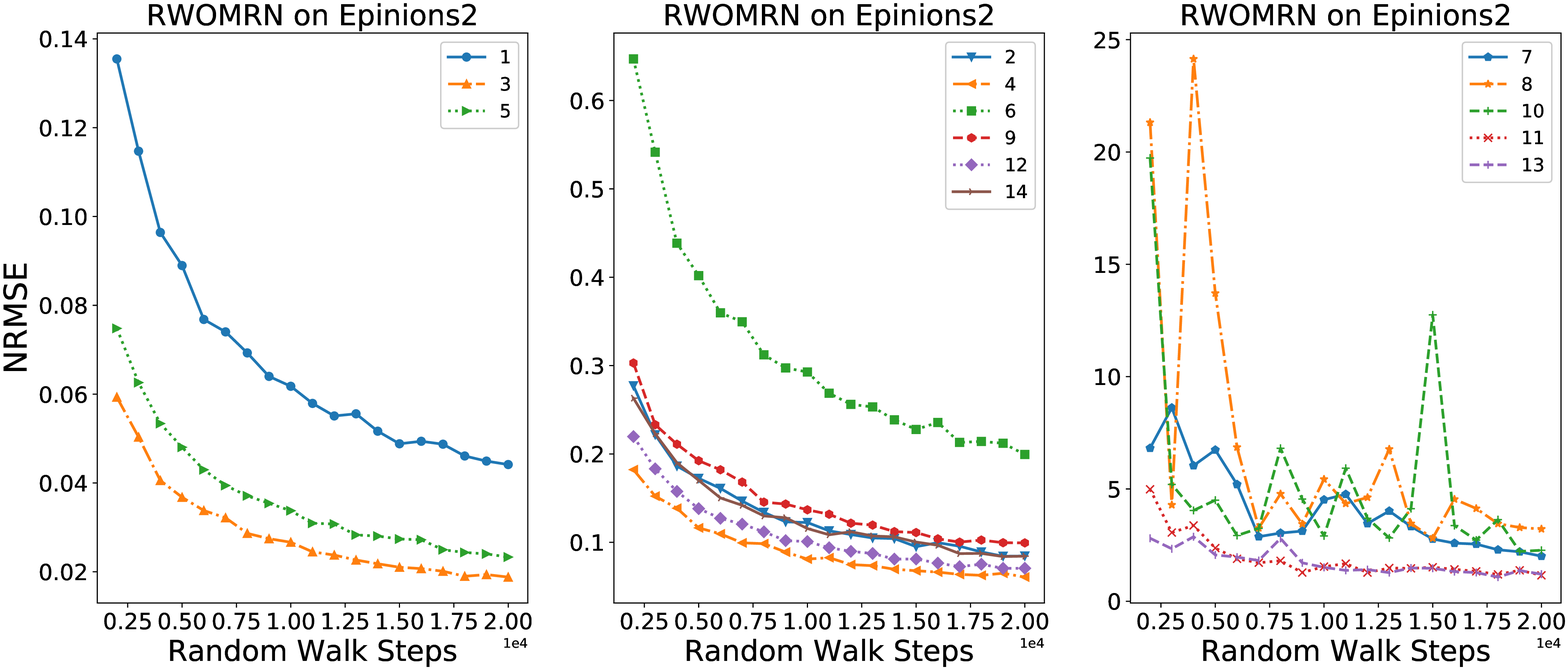}
\caption{RWOMRN on Epinions2}
\label{RWOMRN on Epinions2}
\end{minipage}
\end{figure}

We select two representative algorithms, \textsf{RWNbN} and \textsf{RWMiX}, for analysing the convergence of sampling. 
Fig. \ref{RWNbN on Epinions1} illustrates the NRMSE of different graphlets with  \textsf{RWNbN} when the number of 
sampling steps increases. For most of the graphlets, the NRMSE reduces quickly in the beginning, and tends to converge 
when the sampling step reaches 20$k$. The graphlets in the set $\{1, 3, 5\}$ see the NRMSE at the order of 
$10^{-2}$, and those in the set $\{2, 4, 6, 9, 12, 14\}$ have the NRMSE at the order of $10^{-1}$. 
The $8^{th}$ and $10^{th}$ graphlets contain two red edges that are ``non-overlapping''
with the blue ones so that they are difficult to be sampled under the red layer restriction. At the same time, those two are two of the rarest graphlets in this dataset, that increases the difficulty to sample them. 
Consequently, the NRMSE of the $8^{th}$ and $10^{th}$ graphlets are high, indicating both large errors and large variance. 
Fig. \ref{RWMix on Epinions1} illustrates the NRMSE of different graphlets with \textsf{RWMix}. 
We observe the similar trends as those of \textsf{RWNbN}. Since \textsf{RWMiX} opportunistically samples more red nodes than 
\textsf{RWNbN}, the estimation on the graphlets with more red nodes are expected to be better. 
For instance, the NRMSE of the $4^{th}$ graphlet in \textsf{RWMiX} is obviously lower than that in \textsf{RWNbN}, and 
the NRMSE of the $8^{th}$ and $10^{th}$ graphlets is also slightly lower. 
We further evaluate the MSE of the proposed algorithms on Epinions2 network in Fig. \ref{MRE on Epinions2}. 
The NRMSEs of \textsf{RWEbE} and \textsf{RWOMRN} are shown in Fig. \ref{RWEbE on Epinions2} and Fig. \ref{RWOMRN on Epinions2}, respectively. The experimental results validate the accuracy and convergence of our sampling algorithms in general except for a few extremely rare graphlets. For example, the NRMSE curve of $8^{th}$ graphlet, the rarest one, has the 
highest variance. Whereas the estimation of more frequently graphlets in the set $\{1, 3, 5\}$ is highly accurate and fast convergent. This implies that sampling rare graphlets is even more challenging in multi-layer and restricted graphs. 

\begin{figure}
\begin{minipage}[t]{0.5\textwidth}
\centering
\includegraphics[width=3.3in,height=1.5in]{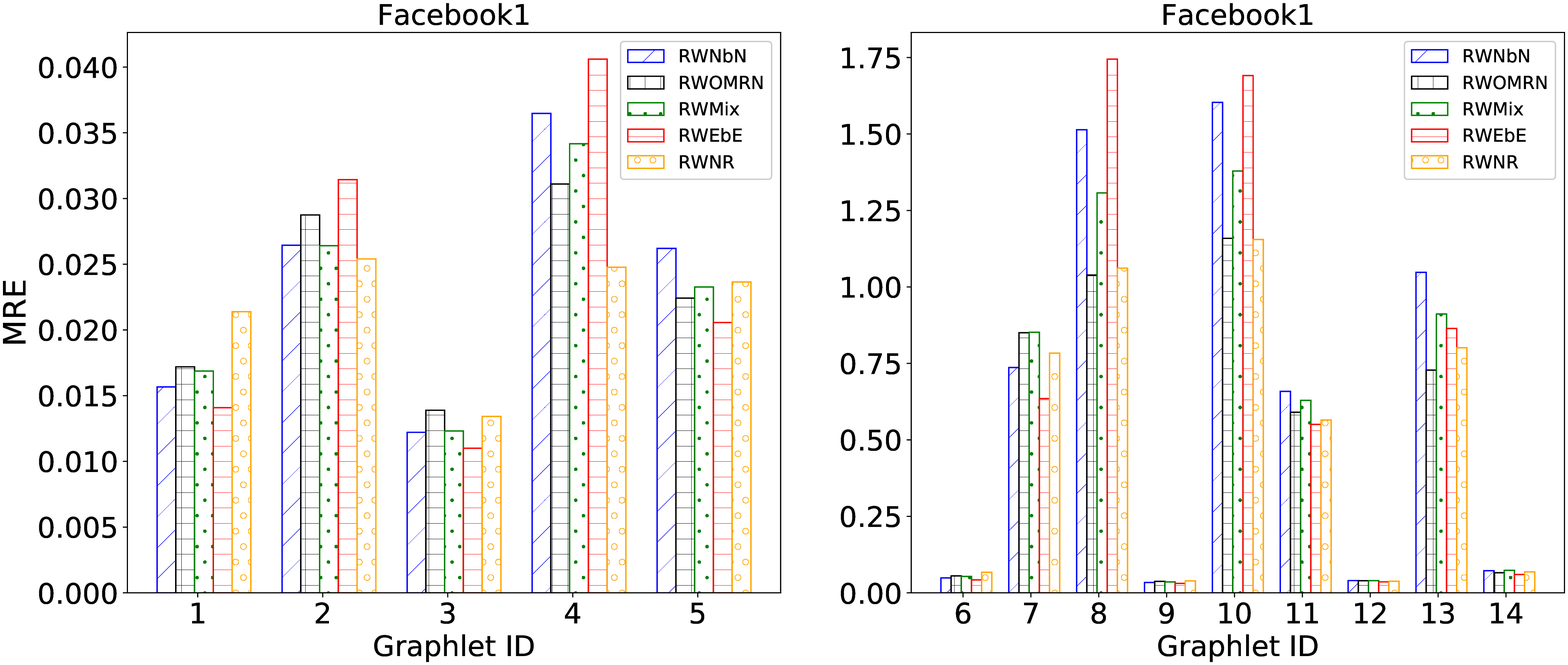}
\caption{MRE on Facebook1}
\label{MRE on Facebook1}
\end{minipage}%
\begin{minipage}[t]{0.5\textwidth}
\centering
\includegraphics[width=3.3in,height=1.5in]{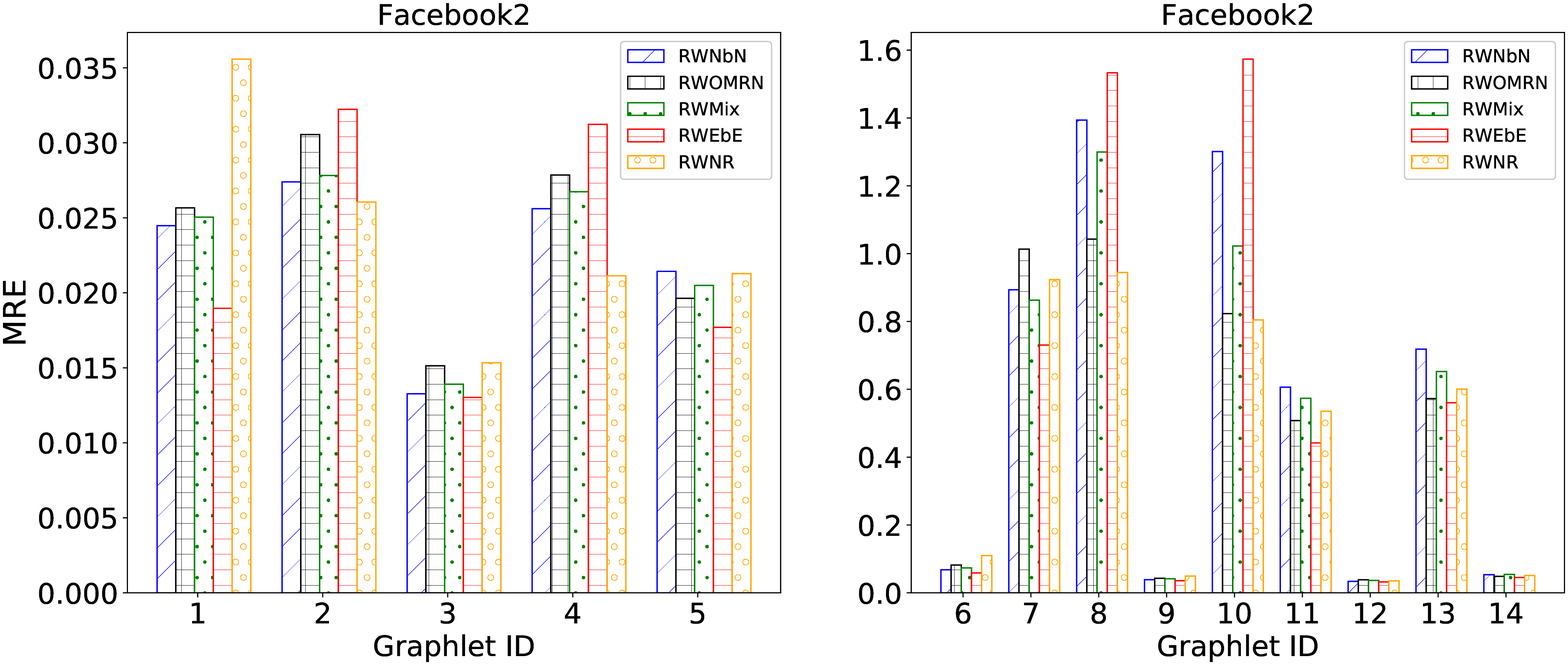}
\caption{MRE on Facebook2}
\label{MRE on Facebook2}
\end{minipage}
\end{figure}

\begin{figure}
\begin{minipage}[t]{0.5\textwidth}
\centering
\includegraphics[width=3.3in,height=1.5in]{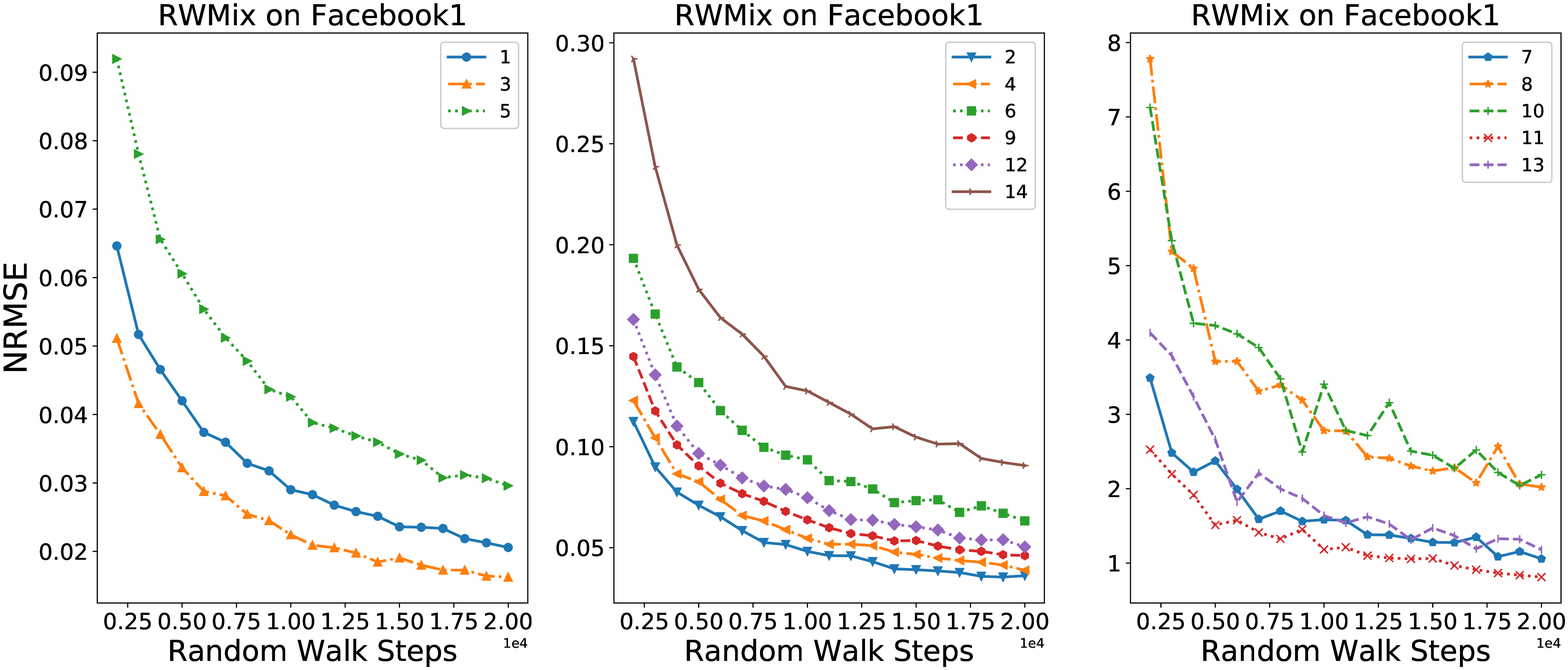}
\caption{RWMix on Facebook1}
\label{RWMix on Facebook1}
\end{minipage}%
\begin{minipage}[t]{0.5\textwidth}
\centering
\includegraphics[width=3.3in,height=1.5in]{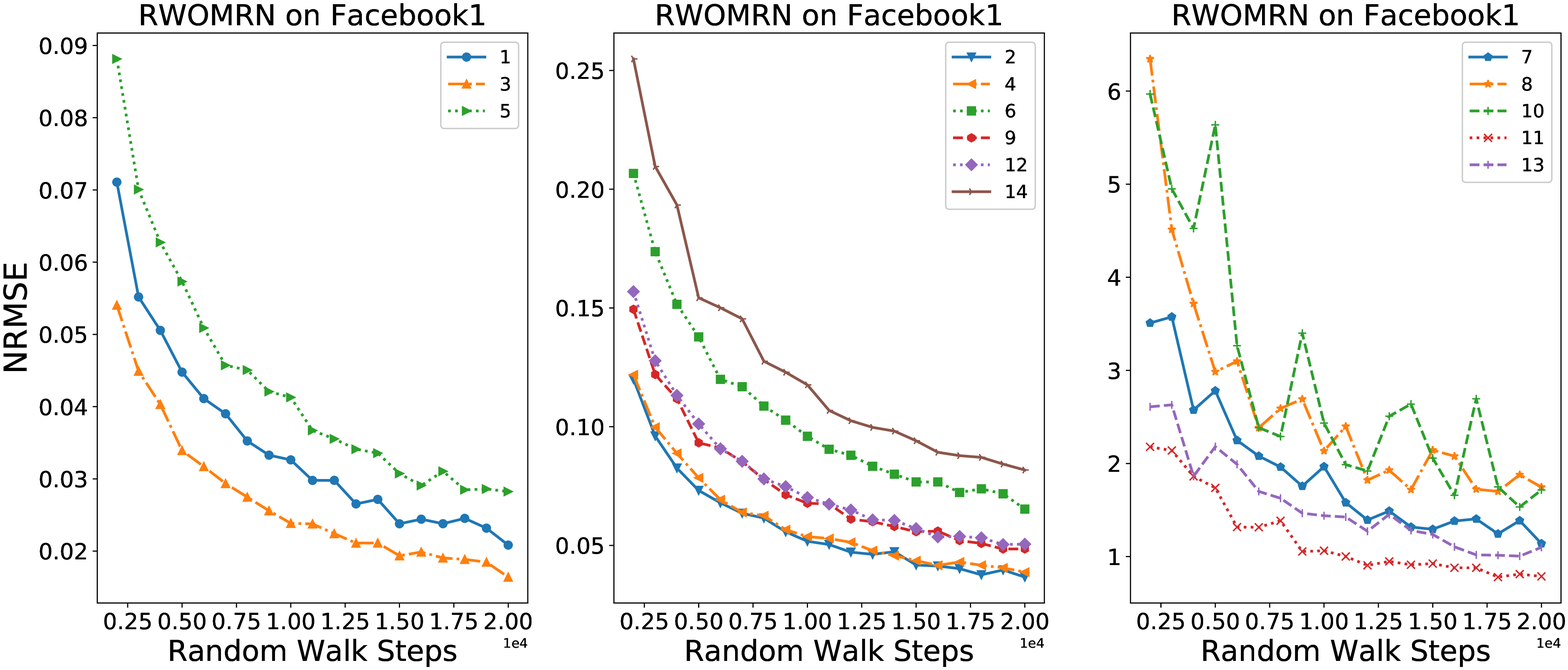}
\caption{RWOMRN on Facebook1}
\label{RWOMRN on Facebook1}
\end{minipage}
\end{figure}

Fig. \ref{MRE on Facebook1} and \ref{MRE on Facebook2} show the MRE of the proposed algorithms with Facebook graphs, 
i.e. Facebook1 and Facebook2. Similar patterns regarding the MRE are observed as those of the Epinions graphs, whereas 
the accuracy of Facebook graphs is better than the Epinions graphs. Fig. \ref{RWMix on Facebook1} and 
\ref{RWOMRN on Facebook1} demonstrate the NRMSE of all the graphlets as the sampling step increases from $2k$ to 20$k$. 
The NRMSE of Facebook graphs is smaller than that of Epinion graphs for each graphlet, and the NRMSE curves of the rare 
graphlets are more smooth, indicating smaller variances in different sampling rounds. The underlying reason is that 
the rare graphlets in the Facebook graphs have relatively high concentration than their counterparts in the Epinions graphs.  

At last, we summarize our answers to the raised questions. 
First, the network structure has a great influence on the sampling accuracy. However, such an influence is not exerted through the type of each graph or the 
way of interconnecting two graphs, but the sparsity of graphlets. 
Second, our proposed algorithm accurately estimates the concentration of the graphlets whose ground-truth percentages are above $10^{-4}$, with a MRE below 0.04. 
They achieve the comparable performance to the benchmark algorithm without random walk restrictions. 
Third, \textsf{RWOMRN} and \textsf{RWMix} usually sample more red nodes, thus 
leading to a relatively higher accuracy for the graphlets with more red nodes.

\section{Conclusion}\label{sec:conclusion}

In this paper, we explore the random walk based sampling on an interconnected two-layer network in which one layer allows random walk and the other layer only permits the one-hop node or edge sampling. 
We present a suite of 3-node graphlets for two-layer multiplex networks, and propose a novel joint 
random work and node sampling approach to perform unbiased estimation of graphlet concentration. 
An analytic bound on the random work steps is derived to achieve asymptotic convergence. 
We observe the inherent tradeoff in the two-layer network with a fixed amount of random walk steps. The concentration of the graphlets with more nodes on one layer can be better estimated when more sampling steps is assigned to this layer. 
We further present two variants to balance the tradeoff between the accuracies of different graphlets. 
Experimental results on the real-world and synthetic graphs manifest that the proposed algorithms can 
accurately estimate the two-layer graphlet concentration, and the level of accuracy is comparable to the 
random walk without layer restriction.



\begin{thebibliography}{0}

\bibitem{roadnetwork}
J.~W. Godfrey, ``The mechanism of a road network,'' \emph{Traffic Engineering
  and Control}, vol.~8, no.~8, 1969.

\bibitem{roadnetwork2}
G.~M. Coclite, M.~Garavello, and B.~Piccoli, ``Traffic flow on a road
  network,'' \emph{SIAM Journal on Mathematical Analysis}, vol.~36, no.~6, pp.
  1862--1886, 2005.

\bibitem{socialnetwork}
J.~Scott, ``Social network analysis,'' \emph{Sociology}, vol.~22, no.~1, pp.
  109--127, 1988.

\bibitem{socialnetwork2}
S.~P. Borgatti, A.~Mehra, D.~J. Brass, and G.~Labianca, ``Network analysis in
  the social sciences,'' \emph{Science}, vol. 323, no. 5916, pp. 892--895,
  2009.

\bibitem{biologicalnetwork}
N.~Pr{\v{z}}ulj, ``Biological network comparison using graphlet degree
  distribution,'' \emph{Bioinformatics}, vol.~23, no.~2, pp. e177--e183, 2007.

\bibitem{nodeclustering}
O.~Younis, M.~Krunz, and S.~Ramasubramanian, ``Node clustering in wireless
  sensor networks: recent developments and deployment challenges,'' \emph{IEEE
  network}, vol.~20, no.~3, pp. 20--25, 2006.

\bibitem{networkmodularity}
J.~M. Hofman and C.~H. Wiggins, ``Bayesian approach to network modularity,''
  \emph{Physical review letters}, vol. 100, no.~25, p. 258701, 2008.

\bibitem{heider1958psychology}
F.~Heider, \emph{The psychology of interpersonal relations}.\hskip 1em plus
  0.5em minus 0.4em\relax Psychology Press, 1958.

\bibitem{kunegis2009slashdot}
J.~Kunegis, A.~Lommatzsch, and C.~Bauckhage, ``The slashdot zoo: mining a
  social network with negative edges,'' in \emph{International Conference on
  World Wide Web}.\hskip 1em plus 0.5em minus 0.4em\relax ACM, 2009, pp.
  741--750.

\bibitem{juszczyszyn2011link}
K.~Juszczyszyn, K.~Musial, and M.~Budka, ``Link prediction based on subgraph
  evolution in dynamic social networks,'' in \emph{International Conference on
  Social Computing}.\hskip 1em plus 0.5em minus 0.4em\relax IEEE, 2011, pp.
  27--34.

\bibitem{zhang2015diffusion}
J.~Zhang, Z.~Fang, W.~Chen, and J.~Tang, ``Diffusion of “following” links
  in microblogging networks,'' \emph{IEEE Transactions on Knowledge and Data
  Engineering}, vol.~27, no.~8, pp. 2093--2106, 2015.

\bibitem{becchetti2008efficient}
L.~Becchetti, P.~Boldi, C.~Castillo, and A.~Gionis, ``Efficient semi-streaming
  algorithms for local triangle counting in massive graphs,'' in
  \emph{Proceedings of the 14th ACM SIGKDD international conference on
  Knowledge discovery and data mining}, 2008, pp. 16--24.

\bibitem{memorybased1}
C.~Seshadhri, A.~Pinar, and T.~G. Kolda, ``Triadic measures on graphs: The
  power of wedge sampling,'' in \emph{International Conference on Data
  Mining}.\hskip 1em plus 0.5em minus 0.4em\relax SIAM, 2013, pp. 10--18.

\bibitem{memorybased2}
M.~Jha, C.~Seshadhri, and A.~Pinar, ``Path sampling: A fast and provable method
  for estimating 4-vertex subgraph counts,'' in \emph{International Conference
  on World Wide Web}.\hskip 1em plus 0.5em minus 0.4em\relax International
  World Wide Web Conferences, 2015, pp. 495--505.

\bibitem{memorybased3}
P.~Wang, J.~Tao, J.~Zhao, and X.~Guan, ``Moss: A scalable tool for efficiently
  sampling and counting 4-and 5-node graphlets,'' \emph{arXiv preprint
  arXiv:1509.08089}, 2015.

\bibitem{memorybased4}
M.~Rahman, M.~Bhuiyan, and M.~A. Hasan, ``Graft: An approximate graphlet
  counting algorithm for large graph analysis,'' in \emph{International
  Conference on Information and Knowledge Management}.\hskip 1em plus 0.5em
  minus 0.4em\relax ACM, 2012, pp. 1467--1471.

\bibitem{streaming1}
M.~Jha, C.~Seshadhri, and A.~Pinar, ``A space efficient streaming algorithm for
  triangle counting using the birthday paradox,'' in \emph{International
  Conference on Knowledge Discovery and Data Mining}.\hskip 1em plus 0.5em
  minus 0.4em\relax ACM, 2013, pp. 589--597.

\bibitem{streaming2}
N.~K. Ahmed, N.~Duffield, T.~L. Willke, and R.~A. Rossi, ``On sampling from
  massive graph streams,'' \emph{International Conference on Very Large Data
  Bases}, vol.~10, no.~11, pp. 1430--1441, 2017.

\bibitem{restrict-cluster}
L.~Katzir and S.~J. Hardiman, ``Estimating clustering coefficients and size of
  social networks via random walk,'' \emph{ACM Transactions on the Web},
  vol.~9, no.~4, p.~19, 2015.

\bibitem{restrict-graphlet}
M.~A. Bhuiyan, M.~Rahman, M.~Rahman, and A.~H. Mohammad, ``Guise: Uniform
  sampling of graphlets for large graph analysis,'' in \emph{International
  Conference on Data Mining}.\hskip 1em plus 0.5em minus 0.4em\relax IEEE,
  2012, pp. 91--100.

\bibitem{restrict}
P.~Wang, J.~Lui, B.~Ribeiro, D.~Towsley, J.~Zhao, and X.~Guan, ``Efficiently
  estimating motif statistics of large networks,'' \emph{ACM Transactions on
  Knowledge Discovery from Data}, vol.~9, no.~2, p.~8, 2014.

\bibitem{restrict-chen}
X.~Chen, Y.~Li, P.~Wang, and J.~Lui, ``A general framework for estimating
  graphlet statistics via random walk,'' \emph{International Conference on Very
  Large Data Bases}, vol.~10, no.~3, pp. 253--264, 2016.

\bibitem{reservoirsample}
J.~S. Vitter, ``Random sampling with a reservoir,'' \emph{ACM Transactions on
  Mathematical Software}, vol.~11, no.~1, pp. 37--57, 1985.

\bibitem{ahmed2017sampling}
N.~K. Ahmed, N.~Duffield, T.~Willke, and R.~A. Rossi, ``On sampling from
  massive graph streams,'' \emph{arXiv preprint arXiv:1703.02625}, 2017.

\bibitem{randomwalk}
J.~D. Noh and H.~Rieger, ``Random walks on complex networks,'' \emph{Physical
  review letters}, vol.~92, no.~11, p. 118701, 2004.

\bibitem{gong2019exploring}
Q.~Gong, Y.~Chen, X.~Yu, C.~Xu, Z.~Guo, Y.~Xiao, F.~B. Abdesslem, X.~Wang, and
  P.~Hui, ``Exploring the power of social hub services,'' \emph{World Wide
  Web}, vol.~22, no.~6, pp. 2825--2852, 2019.

\bibitem{ahmed2015efficient}
N.~K. Ahmed, J.~Neville, R.~A. Rossi, and N.~Duffield, ``Efficient graphlet
  counting for large networks,'' in \emph{International Conference on Data
  Mining}.\hskip 1em plus 0.5em minus 0.4em\relax IEEE, 2015, pp. 1--10.

\bibitem{hovcevar2014combinatorial}
T.~Ho{\v{c}}evar and J.~Dem{\v{s}}ar, ``A combinatorial approach to graphlet
  counting,'' \emph{Bioinformatics}, vol.~30, no.~4, pp. 559--565, 2014.

\bibitem{suri2011counting}
S.~Suri and S.~Vassilvitskii, ``Counting triangles and the curse of the last
  reducer,'' in \emph{International Conference on World Wide Web}, 2011, pp.
  607--614.

\bibitem{streaminggraph}
N.~K. Ahmed, J.~Neville, and R.~Kompella, ``Network sampling: From static to
  streaming graphs,'' \emph{ACM Transactions on Knowledge Discovery from Data},
  vol.~8, no.~2, pp. 1--56, 2013.

\bibitem{restrict-other}
C.~H. Lee, X.~Xu, and D.~Y. Eun, ``Beyond random walk and metropolis-hastings
  samplers: why you should not backtrack for unbiased graph sampling,'' in
  \emph{SIGMETRICS}, vol.~40, no.~1.\hskip 1em plus 0.5em minus 0.4em\relax
  ACM, 2012, pp. 319--330.

\bibitem{restric-other2}
R.~H. Li, J.~X. Yu, L.~Qin, R.~Mao, and T.~Jin, ``On random walk based graph
  sampling,'' in \emph{International Conference on Data Engineering}.\hskip 1em
  plus 0.5em minus 0.4em\relax IEEE, 2015, pp. 927--938.

\bibitem{gjoka2011multigraph}
M.~Gjoka, C.~T. Butts, M.~Kurant, and A.~Markopoulou, ``Multigraph sampling of
  online social networks,'' \emph{IEEE Journal on Selected Areas in
  Communications}, vol.~29, no.~9, pp. 1893--1905, 2011.

\bibitem{li2011sampling}
J.~Y. Li and M.~Y. Yeh, ``On sampling type distribution from heterogeneous
  social networks,'' in \emph{Pacific-Asia Conference on Knowledge Discovery
  and Data Mining}.\hskip 1em plus 0.5em minus 0.4em\relax Springer, 2011, pp.
  111--122.

\bibitem{huang2015triadic}
H.~Huang, J.~Tang, L.~Liu, J.~Luo, and X.~Fu, ``Triadic closure pattern
  analysis and prediction in social networks,'' \emph{IEEE Transactions on
  Knowledge and Data Engineering}, vol.~27, no.~12, pp. 3374--3389, 2015.

\bibitem{dong2012link}
Y.~Dong, J.~Tang, S.~Wu, J.~Tian, N.~V. Chawla, J.~Rao, and H.~Cao, ``Link
  prediction and recommendation across heterogeneous social networks,'' in
  \emph{2012 IEEE 12th International conference on data mining}.\hskip 1em plus
  0.5em minus 0.4em\relax IEEE, 2012, pp. 181--190.

\bibitem{klimek2013triadic}
P.~Klimek and S.~Thurner, ``Triadic closure dynamics drives scaling laws in
  social multiplex networks,'' \emph{New Journal of Physics}, vol.~15, no.~6,
  p. 063008, 2013.

\bibitem{Geyer2005MarkovCM}
C.~J. Geyer, ``Markov chain monte carlo lecture notes,'' 2005.

\bibitem{Chung2012ChernoffHoeffdingBF}
K.-M. Chung, H.~Lam, Z.~Liu, and M.~Mitzenmacher, ``Chernoff-hoeffding bounds
  for markov chains: Generalized and simplified,'' in \emph{STACS}, 2012.

\end{thebibliography}

\newpage
\section*{Appendix}

\subsection*{A:Stationary Distribution}
We have the following two equations:
\begin{equation}
\label{eq1}
\pi \times P = \pi
\end{equation}
\begin{equation}
\label{eq2}
\sum \pi = 1
\end{equation}
The set of all the blue neighbors of $X_m$ is denoted by $B(X_m)$ and the set of all the red neighbors of $X_m$ is denoted by $R(X_m)$.
According to equation (\ref{eq1}), we have:
\begin{equation}
\label{eq3}
\pi(X_m,X_{m+1},Y_{m+2}) = \sum_{a \in B(X_m)}\pi(a,X_m,X_{m+1}) \times \frac{1}{r_{X_{m+1}}+b_{X_{m+1}}} \\
\end{equation}

\begin{equation}
\begin{split}
\label{eq4}
\pi(X_m,X_{m+1},X_{m+2}) = \sum_{a \in B(X_m)}\pi(a,X_m,X_{m+1}) \times \frac{1}{r_{X_{m+1}}+b_{X_{m+1}}}\\
+ \sum_{b \in R(X_{m+1})}\pi(X_m,X_{m+1},b) \times \frac{1}{b_{X_{m+1}}}
\end{split}
\end{equation}

One can check that the following solution satisfies equation (\ref{eq3}) and (\ref{eq4}), where $M$ is a unknown constant.
\[ \begin{cases}
\pi(X_m,X_{m+1},Y_{m+2}) =  \frac{1}{M(r_{X_{m+1}}+b_{X_{m+1}})}\\
\pi(X_m,X_{m+1},X_{m+2}) = \frac{1}{M b_{X_{m+1}}}
\end{cases} \]\\

Now, we can use equation (\ref{eq2}) to solve the unknown constant $M$.

According to equation (\ref{eq2}), we have:
\begin{equation}
\begin{split}
\sum_{v \in \mathcal{V}_B} \sum_{a_1 \in B(v), a_2 \in B(v)} \pi(a_1,v,a_2) \\
+ \sum_{v \in \mathcal{V}_B} \sum_{a_1 \in B(v), b_1 \in R(v)}\pi(a_1,v,b_1) \\
=\sum_{v \in \mathcal{V}_B} {b_v}^2 \frac{1}{M b_v} +  \sum_{v \in \mathcal{V}_B} b_v r_v \frac{1}{M (b_v + r_v)}\\
= \sum_{v \in \mathcal{V}_B}\frac{1}{M}(b_v + \frac{b_v r_v}{b_v + r_v}) = 1
\end{split}
\end{equation}

Which means:
\begin{equation}
M =\sum_{v \in \mathcal{V}_B}(b_v + \frac{b_v r_v}{b_v + r_v}) = 2 |\mathcal{E}_B| + \sum_{v \in \mathcal{V}_B}\frac{b_v r_v}{b_v + r_v}
\end{equation}
Finally, we have the stationary distribution:
\[ \begin{cases}
\pi(X_m,X_{m+1},Y_{m+2}) =  \frac{1}{M(r_{X_{m+1}}+b_{X_{m+1}})}\\
\pi(X_m,X_{m+1},X_{m+2}) = \frac{1}{M b_{X_{m+1}}}
\end{cases} \]\\

It can be checked that the stationary distribution above is the solution of equations (\ref{eq1}) and (\ref{eq2}).

\subsection*{B:Error Bound}
Here is the proof of Theorem \ref{ErrorBound}. It needs several steps.

\begin{lemma}
\label{lemma1}
$\forall$ $0<\delta<1,$ $\exists$ constant $\xi$, such that, when $\forall$ n $\geq$ $\xi \frac{H}{\alpha_i |C_i|}\frac{\tau}{\epsilon^2}(\ln \frac{||\phi||_{\pi}}{\delta})$, we have 
\begin{equation}
Pr (|\frac{\hat {|C_i|}}{|C_i|} - 1| > \frac{\epsilon}{3}) < \frac{\delta}{2}.
\end{equation}
\end{lemma}
Proof: Define $f_i(S) = g_i(S)/(\pi(S)H)$, where $f_i \in [0,1]$. Let 
\begin{equation*}
\mu_i = \mathbb{E}_{\pi}[f_i] = \sum_{S \in W}f_i(S)\pi(S)
\end{equation*}
Which is
\begin{equation}
\sum_{S \in W}g_i(S)/H = \alpha_i |C_i|/H.
\end{equation}
If we have n valid states by random walking with sampling, which are $\{S_j\}_{j=1}^n$, then $\hat {|C_i|} = \frac{1}{n}\sum_{j=1}^n\frac{g_i(S_j)}{\alpha_i \pi(S_j)}$.
\\Define $Z = \sum_{j=1}^n f(S_j) = \sum_{j=1}^n \frac{g_i(S_j)}{\pi(S_j)H}$. According to Theorem \ref{theorem:errorbound}, we have
\[Pr(|\frac{1}{n}\sum_{j=1}^n\frac{g_j(S_j)}{\pi(S_j)H} - \frac{\alpha_i |C_i|}{H}| > \frac{\epsilon}{3} \frac{\alpha_i |C_i|}{H}) \leq c||\phi||_{\pi}e^{\frac{-\epsilon^2 \mu_i n}{648\tau}}\]where c is a constant.
By simplifying, we have \[Pr(|\frac{\hat {|C_i|}}{|C_i|} - 1| > \frac{\epsilon}{3}) \leq c||\phi||_{\pi}e^{\frac{-\epsilon^2 \mu_i n}{648\tau}}\]Assuming that $c||\phi||_{\pi}e^{\frac{-\epsilon^2 \mu_i n}{648\tau}} \leq \frac{\delta}{2} $, we can obtain $n \geq \frac{648\tau}{\epsilon^2 \mu_i}\ln \frac{2c||\phi||_{\pi}}{\delta}$. That means $\exists$ constant  $\xi$, such that, when $n \geq \xi \frac{H}{\alpha_i |C_i|}\frac{\tau}{\epsilon^2}\ln \frac{||\phi||_{\pi}}{\delta}$, we have:
\begin{equation}
Pr(|\frac{\hat {|C_i|}}{|C_i|} - 1| \leq \frac{\delta}{2}).
\end{equation}\\

\begin{lemma}
\label{lemma2}
$\forall 0 < \delta < 1$, $\exists$ constant $\xi$, such that, when $n \geq \xi \frac{H}{\alpha_{min} |C|} \frac{\tau}{\epsilon^2}\ln \frac{||\phi||_{\pi}}{\delta}$, we have 
\begin{equation}
Pr(|\frac{\hat {|C|}}{|C|} - 1| > \frac{\epsilon}{3}) < \frac{\delta}{2}.
\end{equation}
\end{lemma}
Proof: Define 
\begin{equation*}
f(S) = \frac{\mathbb I\{|S| = 3\}}{\sum_{i=1}^{14}\alpha_i g_i(S) \pi(S)} \frac{\alpha_{min}}{H}
\end{equation*}
Then
\begin{equation} 
\mu = \mathbb{E}_{\pi}[f] = \frac{\alpha_{min}}{H}\sum_{S \in W}\frac{\mathbb I\{|S| = 3\}}{\sum_{i=1}^{14}\alpha_i g_i(S)}
\end{equation}
\\Because $\alpha_i |C_i| = \sum_{S \in W}g_i(S)$,
\begin{equation}
|C| = \sum_{i=1}^{14}|C_i| = \sum_{i=1}^{14}\sum_{S \in w}\frac{g_i(S)}{\alpha_i}
\end{equation}
we can find that 
\begin{equation*}
\sum_{i=1}^{14}\sum_{S \in W}\frac{g_i(S)}{\alpha_i} = \sum_{S \in W}\frac{\mathbb I\{|S| = 3\}}{\sum_{i=1}^{14}\alpha_i g_i(S)}
\end{equation*}
So, we have $\mathbb{E}_{\pi}[f] = \frac{\alpha_{min}|C|}{H}$.
\\Define $Z = \sum_{j=1}^n f(S_j)$. According to  Theorem \ref{theorem:errorbound}, we have 
\begin{eqnarray*}
&&Pr(|\frac{1}{n}\sum_{j=1}^n \frac{\mathbb I\{|S| = 3\}}{\sum_{i=1}^{14}\alpha_i g_i(S_j)\pi(S_j)}\frac{\alpha_{min}}{H} - \frac{\alpha_{min}|C|}{H}| > \frac{\epsilon}{3}) \\
&&\leq c||\phi||_{\pi}e^{\frac{- \epsilon^2 \mu n}{648\tau}}
\end{eqnarray*}
We can find that 
\begin{equation*}
\frac{1}{n}\sum_{j=1}^n \frac{\mathbb I\{|S| = 3\}}{\sum_{i=1}^{14}\alpha_i g_i(S_j)\pi(S_j)} = \frac{1}{n}\sum_{j=1}^n \sum_{i=1}^{14} \frac{g_i(S_j)}{\alpha_i \pi(S_j)} = \hat {|C|}
\end{equation*}
So, 
\begin{equation*}
Pr(|\frac{\hat {|C|}}{|C|} - 1| > \frac{\epsilon}{3}) \leq c||\phi||_{\pi}e^{\frac{-\epsilon^2 \mu n}{648\tau}}.
\end{equation*} 
Assuming that $c||\phi||_{\pi}e^{\frac{-\epsilon^2 \mu n}{648\tau}} \leq \frac{\delta}{2} $, we can obtain $n \geq \frac{648\tau}{\epsilon^2 \mu}\ln \frac{2c||\phi||_{\pi}}{\delta}$. That means $\exists$ constant  $\xi$, such that, when $n \geq \xi \frac{H}{\alpha_{min} |C|}\frac{\tau}{\epsilon^2}\ln \frac{||\phi||_{\pi}}{\delta}$, we have 
\begin{equation}
Pr(|\frac{\hat {|C|}}{|C|}| > \frac{\epsilon}{3}) < \frac{\delta}{2}
\end{equation}\\

\begin{theorem}
\label{ErrorBound'}
$\forall$ $0 < \delta < 1, \exists$ constant $\xi$, such that, when $n \geq \xi \frac{H}{\Lambda} \frac{\tau}{\epsilon^2}\ln \frac{||\phi||_{\pi}}{\delta}$, we have 
\begin{equation}
Pr((1-\epsilon)d_i \leq \hat d_i \leq (1+\epsilon)d_i) > 1-\delta.
\end{equation}
\end{theorem}
Proof: If $n$ satisfies the condition in this Theorem, then it must satisfy the conditions in Lemma\ref{lemma1} and Lemma \ref{lemma2}. Let $A_1$ denote the event that $|\frac{\hat {|C_i|}}{|C_i|} - 1| \leq \frac{\epsilon}{3}$, and $A_2$ denote the event that $|\frac{\hat {|C|}}{|C|} - 1| \leq \frac{\epsilon}{3}$. We  have  $Pr(A_1) > 1 - \frac{\delta}{2}$, and $Pr(A_2) > 1 - \frac{\delta}{2}$. If $A_1$ and $A_2$ happen, then
\begin{equation}
(1-\epsilon)d_i \leq \frac{(1-\frac{\epsilon}{3})|C_i|}{(1+\frac{\epsilon}{3})|C|} \leq \frac{\hat {|C_i|}}{\hat {|C|}} \leq \frac{(1+\frac{\epsilon}{3})|C_i|}{(1-\frac{\epsilon}{3})|C|} \leq (1+\epsilon)d_i
\end{equation}
Let F denote the event that $(1-\epsilon)d_i \leq \hat d_i \leq (1+\epsilon)d_i$, then $A_1\cap A_2 \subset F$. So, we have 
\begin{equation}
Pr(F) \geq Pr(A_1\cap A_2) \geq Pr(A_1) + Pr(A_2) -1 = 1 - \delta.
\end{equation}

\subsection*{C:Ground truth of datasets}
Here we present the ground truth of all the concentrations of each dataset.

\begin{table}[!htb]
\centering
\caption{Information of data sets }
\label{tab8}
\begin{tabular}{|c|c|c|c|c|c|c|c|c|}
\hline
Graph&$|\mathcal{V}|$&$|\mathcal{E}|$&\#1($10^{-1}$)&\#2($10^{-1}$)&\#3($10^{-1}$)&\#4($10^{-1}$)&\#5($10^{-2}$)&\#6($10^{-4}$)\\ 
\hline
ER1&100K&948K&1.85&3.03&2.39&1.96&7.73&0.0378\\
\hline
ER2&100K&753K&4.14&2.48&1.68&1.27&4.32&0.135\\
\hline
ER3&100K&753K&4.13&2.47&1.69&1.27&4.34&0.122\\
\hline
SW1&100K&757K&1.72&3.19&2.23&2.06&7.25&14\\
\hline
SW2&120K&824K&2.17&3.98&1.57&1.73&4.55&10.8\\
\hline
SW3&100K&651K&3.25&2.61&1.58&1.93&5.62&33.2\\
\hline
BA1&100K&946K&2.83&1.39&3.68&0.899&12&0.415\\
\hline
BA2&120K&1068K&3.58&2.36&2.92&0.503&6.35&0.448\\
\hline
BA3&100K&957K&2.62&3.3&2.15&1.46&4.69&0.327\\
\hline
Epinions1&76K&842K&2.66&0.378&4.51&0.321&19.2&33.5\\
\hline
Epinions2&76K&990K&1.59&0.4&4.32&0.548&29.4&15.7\\
\hline
Facebook1&63K&1532K&2.68&1.49&3.4&0.94&10.7&95.3\\
\hline
Facebook2&63K&1841K&1.47&1.56&3.18&1.69&17.3&41\\
\hline
\end{tabular}
\end{table}

\begin{table}[!htb]
\centering
\caption{Information of data sets (continued) }
\label{tab8}
\begin{tabular}{|c|c|c|c|c|c|c|c|c|}
\hline
Graph&\#7($10^{-5}$)&\#8($10^{-7}$)&\#9($10^{-5}$)&\#10($10^{-7}$)&\#11($10^{-6}$)&\#12($10^{-5}$)&\#13($10^{-6}$)&\#14($10^{-5}$)\\ 
\hline
ER1&0.836&80.6&0.756&52.7&11.5&0.567&3.68&0.0995\\
\hline
ER2&0.857&39.7&0.584&67.1&5.84&0.211&3.48&0.0994\\
\hline
ER3&1.14&38.5&0.77&49.7&5.09&0.186&4.72&0.0745\\
\hline
SW1&0.542&78.8&279&37.8&8.37&176&3.12&37.7\\
\hline
SW2&147&9620&213&4130&1300&184&282&65.2\\
\hline
SW3&0.586&73.7&141&102&7.75&180&5.29&75.6\\
\hline
BA1&1.37&32.5&8.02&20.6&20&5.48&4.86&0.917\\
\hline
BA2&2.43&127&7.26&21.6&16.2&4.4&5.68&1.03\\
\hline
BA3&4.87&385&5.3&249&62&3.28&21&0.769\\
\hline
Epinions1&2.31&16&858&14.2&40.5&730&16.2&207\\
\hline
Epinions2&1.99&26.4&641&32.3&53.3&876&36.8&398\\
\hline
Facebook1&5.5&148&1800&87.9&68.4&1140&22.1&240\\
\hline
Facebook2&4.38&230&1330&257&96.5&1430&53.6&515\\
\hline
\end{tabular}
\end{table}


%


\ifCLASSOPTIONcaptionsoff
  \newpage
\fi

\end{document}